\documentclass[aps,prd,twocolumn,superscriptaddress,nofootinbib]{revtex4-2}
\usepackage{graphicx}
\usepackage{breqn}
\usepackage{mathtools}
\usepackage{amsmath}
\usepackage[dvipsnames]{xcolor}
\usepackage{xcolor}
\usepackage{aas_macros}

\usepackage{tablefootnote} 
\usepackage{tabularx}
\usepackage{array} 

\newcommand{\Blos}{B_{p_{\hatbf{n}}p_{\hatbf{n}}\delta}}
\newcommand{\Plos}{P_{p_{\hatbf{n}}}}
\newcommand{\Bperp}{B_{p_{\perp}p_{\perp}\delta}}
\newcommand{\Pperp}{P_{p_{\perp}}}
\newcommand{\Ak}{A_{\mathrm{kSZ}^2}}
\newcommand{\CTTg}{C_{\ell}^{\mathrm{kSZ}^{2}\times\delta_{g}}}
\newcommand{\lens}{C_{\ell}^{\mathrm{lens}}}
\newcommand{\mnus}{\Sigma m_{\nu}}

\newcommand{\hatbf}[1]{\mathbf{\hat{#1}}}

\newcommand\numberthis{\addtocounter{equation}{1}\tag{\theequation}}

\graphicspath{{./}{figures/}}

\begin{document}


\title{Improved Modeling of the Kinematic Sunyaev-Zel'dovich Projected-Fields signal and its Cosmological Dependence}


\author{Raagini Patki}
\affiliation{Department of Astronomy\char`,{} Cornell University\char`,{} Ithaca\char`,{} NY 14853\char`,{} USA.}
\author{Nicholas Battaglia}
\affiliation{Department of Astronomy\char`,{} Cornell University\char`,{} Ithaca\char`,{} NY 14853\char`,{} USA.}
\author{Simone Ferraro}
\affiliation{Lawrence Berkeley National Laboratory\char`,{} One Cyclotron Road\char`,{} Berkeley\char`,{} CA 94720\char`,{} USA.}
\affiliation{Berkeley Center for Cosmological Physics\char`,{} Department of Physics\char`,{} University of California\char`,{} Berkeley\char`,{} CA 94720\char`,{} USA.}


\begin{abstract}
Over the past decade, the kinematic Sunyaev-Zel'dovich (kSZ) effect has emerged as an observational probe of the distribution of baryons and velocity fields in the late Universe. Of the many ways to detect the kSZ, the `projected-fields kSZ estimator' has the promising feature of not being limited to galaxy samples with accurate redshifts. The current theoretical modeling of this estimator involves an approximate treatment only applicable at small scales. As the measurement fidelity rapidly improves, we find it necessary to move beyond the original treatment and hence derive an \emph{improved} theoretical model for this estimator without these previous approximations. We show that the differences between the predicted signal from the two models are scale-dependent and will be significant for future measurements from the Simons Observatory and CMB-S4 in combination with galaxy data from WISE or the Rubin Observatory, which have high forecasted signal-to-noise ratios ($>100$). Thus, adopting our improved model in future analyses will be important to avoid biases. Equipped with our model, we explore the cosmological dependence of this kSZ signal for future measurements. With a \textit{Planck} prior, residual uncertainty on $\Lambda$CDM parameters leads to $\sim7\%$ marginalized uncertainties on the signal amplitude, compared to a sub-percent level forecasted with a fixed cosmology. To illustrate the potential of this kSZ estimator as a cosmological probe, we forecast initial constraints on $\Lambda$CDM parameters and the sum of neutrino masses, paving the way for jointly fitting both baryonic astrophysics and cosmology in future analyses.
\end{abstract}

\maketitle

\section{\label{sec:intro}Introduction}
Upcoming surveys such as the \textit{Simons Observatory} (SO) \cite{SO} and CMB-S4 \cite{S4} will map the cosmic microwave background (CMB) at arcminute-resolution, allowing the precise measurement of secondary CMB anisotropies which arise due to the large-scale structure (LSS) at low redshifts. The kinematic Sunyaev-Zel'dovich (kSZ) effect \cite{SZ1972, SZ1980, Ostriker1986} is one of the dominant sources of secondaries at arcminute scales. It is the Doppler boosting of CMB photons as they Compton-scatter off of free electrons moving with a non-zero line-of-sight (LOS) velocity. The resulting shift in the observed CMB temperature is proportional to the free electron momentum along the LOS, thus making the kSZ a key probe of both astrophysics (e.g.\,\cite{F16, Battaglia2017, Battaglia2019, Amodeo2020, LaPlante2022}) and cosmology (e.g.\,\cite{Bhattacharya2008, Mueller2014DE, Mueller2014, RoncarelliMoD, Roncarellimnu}). 

The kSZ effect was first detected in 2012 in \textit{Atacama Cosmology Telescope} (ACT) data \cite{Hand2012} using the pairwise momenta of galaxies (e.g. \cite{Ferreira1999}). This method was subsequently used for detections in data from 
\textit{Planck} \cite{Planckpairwise}, the \textit{South Pole Telescope} \cite{SPTpairwise}, and with updated ACT DR5 maps \cite{Calafut2021}. A velocity-weighted stacking approach \cite{Ho2009, Shao} has also been used for measurements in \textit{Planck} \cite{Planck_Hern} and ACT maps \cite{Schaan2016, Schaan2021}. 
While there are no measurements with the recently proposed velocity reconstruction method (e.g.\,\cite{Smith2018}) presently, it is forecasted to become a promising cosmological probe (e.g.\,\cite{Munch, giri}).
A limitation of these kSZ estimators is that they need accurate spectroscopic redshifts, typically available only for smaller galaxy samples. If photometric data is used instead, their signal-to-noise ratios (SNR) decrease sharply \cite{K_Schmidt2013, Flender2016}, despite the increase in the number of galaxies. 

In this paper, we focus on the `projected-fields' estimator for the kSZ \cite{Dore2004, D05}, which has the advantage that it only needs a statistical redshift distribution of the LSS tracers, rather than individual redshifts. This distribution can be obtained from photometric galaxy surveys such as \textit{Wide-field Infrared Survey Explorer} (WISE) \cite{WISE} and the future \textit{Vera C. Rubin Observatory} (VRO) \cite{rubin}, which are less expensive and map larger volumes. First proposed in \cite{Dore2004}, this estimator extracts the kSZ signal that is associated with a given tracer population of the LSS, from the CMB temperature map. Since the LOS velocity of ionized gas (and the corresponding kSZ signal) is equally likely to be positive or negative, a foreground-cleaned and filtered CMB map is first squared in real space, and then cross-correlated with a projected tracer field. Thus, this kSZ${}^2$-LSS cross-correlation is the simplest such non-trivial estimator that does not require 3D information. 

Following the initial formulation by \cite{Dore2004}, \cite{D05} further studied this estimator and suggested an approximate theoretical model for it at small scales (hereafter `D05'). The kSZ${}^2$-LSS cross-correlation was first detected in data from \textit{Planck} and \textit{WMAP} with WISE galaxies \cite{Hill2016, F16}, and then with unWISE galaxies \cite{Kusiak2021}. The measured signal amplitude ($\Ak$) was used to address the `missing baryon' problem by inferring the baryon fraction $f_b$, which is proportional to the free electron density at the corresponding galaxy redshifts. Recently, \cite{Bolliet2022} forecasted future limits on astrophysical parameters governing the baryon density profile with this signal. 

While baryonic effects are largely dominant at small scales, a wider range of scales is affected by the cosmological dependence of the signal. This motivates us to do the \textit{first} derivation of an `improved' theoretical model for the projected-fields estimator that is accurate across these larger scales, while the D05 model used so far is only applicable at small scales. Moreover, we show that there are $\sim$15$\%$ level differences between the signal predicted by our model and the D05 model, which are significant for future CMB surveys that have been forecasted to yield large SNR for the projected-fields estimator \cite{F16, Bolliet2022}. Since these differences are scale-dependent, as well as large \textit{even} at small scales, the resulting change in the shape of the signal would likely impact \emph{any} parameter inference with this estimator.

This improved theoretical model enables us to explore the cosmological dependence of the signal, paving the way for future analyses to jointly fit both astrophysical and cosmological parameters. We study the cosmological sensitivity of the improved signal for the first time, assuming a $\Lambda$CDM cosmology with best-fit \textit{Planck}-2018 parameters \cite{Planck2018} as our baseline fiducial model. Since the kSZ-induced anisotropy is proportional to the LOS velocity of free electrons, it is an unbiased probe of the total underlying matter density and the growth rate of LSS. Massive neutrinos directly impact the clustering and growth rate of LSS (e.g. \cite{Lesgourgues, mnu_whitep}). Therefore, we additionally vary the sum of neutrino masses ($\mnus$) about a fiducial value of 60 meV, as an illustration of the potential sensitivity of kSZ measurements using this estimator to physics beyond the Standard Model.  

The rest of this paper is organized as follows: In Section II, we review the theoretical formalism of the estimator, its existing approximate model, and our `improved' model, providing its rigorous derivation in the Appendix. In Section III, we describe the survey specifications and our numerical implementation. We compare the predicted signal from our model and the D05 model for future measurements in Section IV. In Section V, we present initial Fisher forecasts jointly fitting for $\Lambda$CDM parameters, $\mnus$, $\Ak$, and the galaxy bias, for future SO and CMB-S4 data combined with WISE and a VRO-like experiment. We show the effect of marginalizing over cosmological parameters and the impact of residual foregrounds present in cleaned CMB maps on these forecasts, and discuss future directions. We conclude the paper in Section VI. 

\section{Theoretical Formalism} 
\subsection{The Projected-Fields kSZ estimator} \label{proj}
The kSZ effect leads to a shift in the CMB temperature anisotropies in a direction $\hatbf{n}$ on the sky, $\Theta^{\mathrm{kSZ}}\left( \hatbf{n}\right) \equiv \Delta T^{\mathrm{kSZ}}/T_{\mathrm{CMB}} \left( \hatbf{n}\right)$, (taking c = 1): 
\begin{equation}\label{kSZ}
\Theta^{\mathrm{kSZ}}\left( \hatbf{n}\right)= -\int d\eta \hspace{0.05 cm} g(\eta) \hspace{0.1 cm} \mathbf{p_{e}}\cdot\hatbf{n},     
\end{equation}
where $\eta(z)$ is the comoving distance to redshift $z$, and $\mathbf{p_{e}} = (1 + \delta_{e}) \mathbf{v}_{e}$ is the electron momentum field, with $\delta_{e}$ being the electron overdensity, and $\mathbf{v}_{e}$ the peculiar velocity of electrons. $g(\eta) = e^{-\tau} d\tau/d\eta$ is the visibility function, and $\tau$ is the optical depth to Thomson scattering. Since $d\tau/d\eta = \sigma_{\mathrm{T}}\, n_e/(1+z)$, where $\sigma_{\mathrm{T}}$ is the Thomson scattering cross-section and $n_e$ is the free electron number density,
\begin{equation}
\Theta^{\mathrm{kSZ}}\left( \hatbf{n}\right)= -\sigma_{\mathrm{T}} \int \frac{d\eta}{1+z} e^{-\tau} n_e\left( \hatbf{n}, \eta \right) \mathbf{v_{e}}\left( \hatbf{n}, \eta \right)\cdot\hatbf{n}.     
\end{equation}

In this work, we consider galaxies as a tracer of the LSS. However, the same formalism can be applied equivalently to other tracers of the underlying matter density field, such as quasars, 21-cm fluctuations (e.g.~\cite{Ma2018, LaPlante2022}), or weak lensing convergence (e.g.~\cite{Bolliet2022}). The projected galaxy overdensity field $\delta_{g}(\hatbf{n})$ is:
\begin{equation}\label{delg}
    \delta_{g}(\hatbf{n}) = \int_{0}^{\eta_{\max}} d\eta \hspace{0.07 cm} W^{g}(\eta) \hspace{0.07 cm} \delta_{m}(\eta\hatbf{n}, \eta),
\end{equation}
where $\delta_{m}(\eta\hatbf{n}, \eta)$ is the 3D matter overdensity field. $\eta_{\max}$ is the maximum comoving distance of the galaxy sample, and its projection kernel $W^{g}(\eta) = b_g p_s(\eta)$, where $p_s(\eta) \propto dn/d\eta$ is the redshift distribution of the number of galaxies, normalized to have a unit integral. Here, we assume a linear galaxy bias $b_g$ for simplicity.
 
We apply a Wiener filter $F(\ell)$ to foreground-cleaned CMB blackbody temperature maps in harmonic space, in order to select angular scales dominated by the kSZ contribution \cite{F16, Hill2016} :  
\begin{equation}\label{wiener}
F(\ell) = \frac{C_{\ell}^{\mathrm{kSZ}}}{C_{\ell}^{\mathrm{tot}}}.
\end{equation}
Recently, \cite{Bolliet2022} have proposed a slightly more optimal filter, where $C_{\ell}^{\mathrm{kSZ}}$ in the numerator of Eq.\eqref{wiener} is replaced by $\sqrt{C_{\ell}^{\mathrm{kSZ}}}$. While this is slightly more optimal, here we adopt a classical Wiener filter as in Eq.\eqref{wiener} for comparison with previous works that adopt this choice.

Here, $C_{\ell}^{\mathrm{kSZ}}$ is the theoretical kSZ power spectrum, while $C_{\ell}^{\mathrm{tot}}$ is the total power spectrum, which includes the primary CMB, kSZ, and integrated Sachs-Wolfe (ISW) contributions, detector noise, and residual foregrounds. In addition, telescopes observe the CMB through a finite beam $b(\ell)$, 
\begin{equation} \label{beam}
b(\ell) = \mathrm{exp} \left(-\frac{1}{2}\ell(\ell+1)\frac{\theta_{\mathrm{FWHM}}^{2}}{8\,\mathrm{ln}\,2} \right),     
\end{equation}
where $\theta_{\mathrm{FWHM}}$ is the full width at half maximum in radians, for a Gaussian beam. Therefore, overall, the filtered CMB anisotropies $\Theta_f(\ell)$ are related to the true CMB anisotropies $\Theta(\ell)$ as:
\begin{equation}
\Theta_f(\mathbf{\ell}) = F(\ell)b(\ell)\Theta(\mathbf{\ell}) \equiv f(\ell)\Theta(\mathbf{\ell}),   
\end{equation}
where $f(\ell) \equiv F(\ell)b(\ell)$. Further details about our numerical evaluation of the filters are given in Section \ref{survey}, and they are shown in Figure \ref{fig:fl}.

Free electrons at low-redshift are equally likely to be moving towards or away from us along the LOS. Due to this $\mathbf{v_{e}}\rightarrow-\mathbf{v_{e}}$ parity symmetry, the naive cross-correlation between $\Theta^{\mathrm{kSZ}}$ and $\delta_g$ is expected to vanish at small scales, (while the contribution from linear ISW is detectable at $\ell < 100$ \cite{F16}). Hence, we follow \cite{Dore2004, D05} and \textit{square} the filtered CMB map in real space, before cross-correlating it with $\delta_g$:
\begin{equation}\label{dirac}
\langle \Theta_f^{2}(\mathbf{\ell}) \, \delta_g(\mathbf{\ell^'})\rangle = (2\pi)^{2}\delta_{D}(\mathbf{\ell}+\mathbf{\ell^'}) C_{\ell}^{\mathrm{kSZ}^{2}\times\delta_{g}}.  
\end{equation}

Using the Limber (Kaiser) approximation \cite{Kaiser1992, Limber}, the angular power spectrum of the $\mathrm{kSZ}^2$-galaxy cross-correlation above (i.e. the projected-fields estimator) can be simplified in harmonic space and written as \cite{Dore2004, D05}:
\begin{equation}\label{Cl-def}
    C_{\ell}^{\mathrm{kSZ}^{2}\times\delta_{g}} = \int_{0}^{\eta_{\max}} \frac{d\eta}{\eta^{2}} W^{g}(\eta) g^{2}(\eta) \mathcal{T}\left(j = \frac{\ell}{\eta}, \eta\right),
\end{equation}
where the `triangle power spectrum' $\mathcal{T}$ is 
\begin{equation}\label{triangle}
    \mathcal{T}(j, \eta) =\int \frac{d^{2}\mathbf{q}}{(2\pi)^{2}} f(q\eta) f(|\mathbf{j} + \mathbf{q}|\eta) \, \Blos(\mathbf{q}, -\mathbf{j} - \mathbf{q}, \mathbf{j}).
\end{equation}
Here, the `hybrid bispectrum' $\Blos$ is the three-point correlation in Fourier space between two electron momenta along the LOS ($p_{\hatbf{n}} = \mathbf{p}\cdot\hatbf{n}$), and one fractional overdensity.
$\mathcal{T}(j, \eta)$ is an integral over all (closed) triangles lying in a constant-redshift plane (corresponding to $\eta(z)$) which have one side of length $j$, where the wavevector $j \approx \ell/\eta$ follows from the Limber approximation. $\mathcal{T}$ compresses the rich information contained in $\Blos$ into what is needed to compute $C_{\ell}^{\mathrm{kSZ}^{2}\times\delta_{g}}$.

\subsection{Existing Approximate Model for the Hybrid Bispectrum} \label{D05}
$\Blos$ is the key quantity that must be modeled accurately enough, in order to interpret the measured signals appropriately. The electron momentum field is: $\mathbf{p}(\mathbf{x}) = (1 + \delta(\mathbf{x})) \mathbf{v}(\mathbf{x})$ (we drop the subscript $e$ for clarity), which transforms into Fourier space as: 
$\mathbf{p}(\mathbf{k}) = \mathbf{v}(\mathbf{k}) + \int d^{3}k' \mathbf{v}(\mathbf{k}') \delta(\mathbf{k}-\mathbf{k}')/(2\pi)^3$. In linear theory, velocities $\mathbf{v}(\mathbf{x})$ are purely gradient, and
\begin{equation}\label{lin-vel}
   \mathbf{v}(\mathbf{k}) = i\frac{faH\delta(\mathbf{k})}{k} \hatbf{k},
\end{equation}
where $a, f$, and $H$ are the scale factor, linear growth rate, and Hubble parameter at the corresponding redshift, respectively. In the Limber \cite{Kaiser1992} or `weak coupling' \cite{HuWhite96} approximations, only those $\mathbf{p}(\mathbf{k})$ modes with $\mathbf{k}$ perpendicular to the LOS contribute to the kSZ effect (Eq.~\ref{kSZ}). Therefore, the linear effect from the first term $\mathbf{p} \sim \mathbf{v}$ vanishes \cite{Dore2004}. However, the second term $\mathbf{p} \sim \delta\mathbf{v}$ can have a non-zero ``curl" or transverse component ($\equiv \mathbf{p}_\perp$) that has velocity perpendicular to its wavevector \cite{Ostriker1986}. To capture the full kSZ effect including nonlinear scales, we follow \cite{D05} and only consider the $\delta\mathbf{v}$ term while substituting the non-linear density field in Eq.\eqref{lin-vel} for the velocity. \cite{Zhang2004} first proposed this phenomenologically, and showed that it agrees excellently with their simulations. 

Under the Limber approximation, the power spectrum of the LOS component $p_{\hatbf{n}}$ is $\Plos = (1/2)\Pperp$ \cite{Ma2002, Hu_2000}. Similarly, our hybrid bispectrum $\Blos = (1/2)\Bperp$. The fully-nonlinear improved expression for $\Plos$ was derived in \cite{Ma2002} (see Eq.(6) therein); they show that it simplifies to $\Plos(k) \approx (1/3)v_{\mathrm{rms}}^2 P_{\delta\delta}(k)$ in the high-$k$ (i.e. small scales) limit, where $v_{\mathrm{rms}}^2$ is the 3D velocity dispersion and $P_{\delta\delta}$ is the nonlinear matter power spectrum. The original paper \cite{Dore2004} suggested an analogous \textit{ansatz} in the high-$k_1,k_2,k_3$ limit:
\begin{equation}\label{approx}
    \Blos^{\mathrm{approx}}(\mathbf{k_1}, \mathbf{k_2}, \mathbf{k_3}) \approx \frac{1}{3} v_{\mathrm{rms}}^2 B_{\delta\delta\delta}(\mathbf{k_1}, \mathbf{k_2}, \mathbf{k_3}),
\end{equation}
where $B_{\delta\delta\delta}$ is the nonlinear matter bispectrum. \cite{D05} (D05) provided a rough justification for the application of this \textit{ansatz} for estimating $C_{\ell}^{\mathrm{kSZ}^{2}\times\delta_{g}}$ at small scales (i.e. high-$\ell$); we outline it below. We refer to Eq.\eqref{approx} as the `D05' or `approximate' model for calculating $\CTTg$. While all previous works using this estimator assume this model, it makes some approximations that limit its validity for future surveys; we address them in our \emph{improved} model described in the next subsection. 

By definition, we have (similar to Eq.~\ref{dirac}): 
\begin{align*}
    \langle p_{\hatbf{n}}(\mathbf{k_1}) p_{\hatbf{n}}(\mathbf{k_2}) \delta(\mathbf{k_3}) \rangle = 
    (2\pi)^{3}\delta_{D}(\mathbf{k_1}+\mathbf{k_2}+\mathbf{k_3}) \\ 
    \Blos(\mathbf{k_1}, \mathbf{k_2}, \mathbf{k_3}),  \numberthis \label{Fourier}
\end{align*}
where the Dirac delta function enforces that the three wavevectors must form a closed triangle, and $\langle p_{\hatbf{n}}p_{\hatbf{n}}\delta \rangle = (1/2) \langle p_{\perp}p_{\perp}\delta \rangle$. Since $\mathbf{p} \sim \delta\mathbf{v}$, $\langle p_{\perp}p_{\perp}\delta \rangle$ is a five-point function $\langle \delta\mathbf{v}\delta\mathbf{v}\delta \rangle$, whose Wick contraction consists of $\binom 52 = 10$ different terms that are products of two- and three- point functionals (Table \ref{tab:wicks}), as first explained by D05. The connected five-point term is assumed to be negligible due to weak non-Gaussianity (as demonstrated for $\Plos$ in simulations \cite{Ma2002}), and the `tree-level' expansion is sufficient. Six out of the ten terms turn out to be zero due to the parity of the functionals. 

\begin{table}[!htb]
\centering
\begin{ruledtabular}
\begin{tabular}{|c|c|}
Terms & Geometric scaling \\
\hline
$\langle v^i(\mathbf{k}) v^j(\mathbf{k}^\prime)\rangle\langle \delta(\mathbf{k}_1-\mathbf{k})\delta(\mathbf{k}_2-\mathbf{k}^\prime)\delta(\mathbf{k}_3)\rangle$ & 1 \\
$\langle v^i(\mathbf{k}) \delta(\mathbf{k}_1-\mathbf{k}) \rangle\langle v^j(\mathbf{k}^\prime)\delta(\mathbf{k}_2-\mathbf{k}^\prime)\delta(\mathbf{k}_3)\rangle$ & 0 \\
$\langle v^i(\mathbf{k})\delta(\mathbf{k}_2-\mathbf{k}^\prime)\rangle\langle \delta(\mathbf{k}_1-\mathbf{k})v^j(\mathbf{k}^\prime)\delta(\mathbf{k}_3)\rangle$ & $k/k_2$ \\
$\langle \delta(\mathbf{k}_2-\mathbf{k}^\prime) v^j(\mathbf{k}^\prime)\rangle\langle \delta(\mathbf{k}_1-\mathbf{k})v^i(\mathbf{k})\delta(\mathbf{k}_3)\rangle$ & 0 \\
$\langle\delta(\mathbf{k}_1-\mathbf{k}) v^j(\mathbf{k}^\prime)\rangle\langle  v^i(\mathbf{k})\delta(\mathbf{k}_2-\mathbf{k}^\prime)\delta(\mathbf{k}_3)\rangle$ & $k/k_1$ \\
$\langle v^i(\mathbf{k}) \delta(\mathbf{k}_3)\rangle\langle \delta(\mathbf{k}_1-\mathbf{k})\delta(\mathbf{k}_2-\mathbf{k}^\prime) v^j(\mathbf{k}^\prime)\rangle$ & 0 \\
$\langle \delta(\mathbf{k}_3) v^j(\mathbf{k}^\prime)\rangle\langle \delta(\mathbf{k}_1-\mathbf{k})\delta(\mathbf{k}_2-\mathbf{k}^\prime)v^i(\mathbf{k})\rangle$ & 0 \\
$\langle\delta(\mathbf{k}_1-\mathbf{k})\delta(\mathbf{k}_2-\mathbf{k}^\prime) \rangle\langle v^i(\mathbf{k}) v^j(\mathbf{k}^\prime)\delta(\mathbf{k}_3)\rangle$ & $[-k+(k_1$ or $k_2)]/k_3$ \\

$\langle \delta(\mathbf{k}_2-\mathbf{k}^\prime)\delta(\mathbf{k}_3)\rangle\langle v^i(\mathbf{k}) v^j(\mathbf{k}^\prime) \delta(\mathbf{k}_1-\mathbf{k})\rangle$ & 0 \\
$\langle\delta(\mathbf{k}_1-\mathbf{k})\delta(\mathbf{k}_3)\rangle\langle  v^i(\mathbf{k}) v^j(\mathbf{k}^\prime)\delta(\mathbf{k}_2-\mathbf{k}^\prime)\rangle$ & 0 \\
\end{tabular}
\end{ruledtabular}
\caption{The ten terms in the Wick contraction of $\langle p_{\perp}p_{\perp}\delta \rangle \sim \langle \delta\mathbf{v}\delta\mathbf{v}\delta \rangle$ which contribute to $\Blos(\mathbf{k_1}, \mathbf{k_2}, \mathbf{k_3})$, as first explained in D05 \cite{D05}.}
\label{tab:wicks}
\end{table}

D05 stated the following \textit{ansatz} for the contribution to $\Bperp^{\mathrm{approx}}(\mathbf{k_1}, \mathbf{k_2}, \mathbf{k_3})$ from the leading order (or `usual') term (first row in Table \ref{tab:wicks}) among the four non-zero terms:
\begin{equation}\label{dedeo}
    \int \frac{d^{3}k}{\left( 2\pi \right) ^{3}} [1 - \mu_{1}\mu_{2}] \, P_{vv}(k) B_{\delta\delta\delta }( \mathbf{k_{1}}-\mathbf{k}, \mathbf{k_{2}}+\mathbf{k},\mathbf{k_{3}}),
\end{equation}
where $\mu_{1}\equiv \hatbf{k_1}\cdot\hatbf{k}$ and $\mu_{2}\equiv \hatbf{k_2}\cdot\hatbf{k}$. The kSZ effect is dominant at small scales in the CMB, and thus the filters (e.g.\,Fig.~\ref{fig:fl}) peak at high-$\ell$. D05 then argued that for this small-scale regime of the $C_{\ell}^{\mathrm{kSZ}^{2}\times\delta_{g}}$ estimator, only those triangles with high-$k_1,k_2,k_3$ need to be considered for $\Blos(\mathbf{k_1}, \mathbf{k_2}, \mathbf{k_3})$ in Eq.\eqref{triangle}. In this limit, their \textit{ansatz} above simplifies to: 
\begin{align*}
    \Blos(\mathbf{k_1}, \mathbf{k_2}, \mathbf{k_3}) 
    \approx \frac{1}{2}[1 - \mu_{12}]v_{\mathrm{rms}}^2 B_{\delta\delta\delta}(\mathbf{k_1}, \mathbf{k_2}, \mathbf{k_3}),
\end{align*}
which is close to the \cite{Dore2004} \textit{ansatz} (Eq.~\ref{approx} above), where $\mu_{12}\equiv \hatbf{k_1}\cdot\hatbf{k_2}$. The three other non-zero terms listed in Table \ref{tab:wicks} turn out to have geometric scalings $\propto 1/k_{i}$ (where $i = 1, 2,$ or 3). Since D05 chose to confine to the high-$k_1,k_2,k_3$ limit, they argued that these additional `extra' terms can be neglected. 

\subsection{Our Improved Model}\label{improv}
While the framework and initial basic assumptions for the D05 model are well-founded, they make a series of \textit{further} approximations that we find to be inaccurate. We list these inaccuracies here and describe how our rigorously derived \emph{improved} model addresses and improves upon them:
\begin{itemize}
    \item By definition, the calculation of triangle power spectra $\mathcal{T}(j, \eta)$ given by Eq.\eqref{triangle} is a convolution, in which the shape of the hybrid bispectrum $\Blos(\mathbf{q}, -\mathbf{j} - \mathbf{q}, \mathbf{j})$ gets modified by the filtering product $f(q\eta) f(|\mathbf{j} + \mathbf{q}|\eta)$. Therefore, even if one restricts to the high-$\ell$ regime for $C_{\ell}^{\mathrm{kSZ}^{2}\times\delta_{g}}$, the convolution leads to a mixing of angular scales, and it is \textit{not} sufficient to only consider the high-$k_1,k_2,k_3$ limit. Note that here, $(\mathbf{k_1}, \mathbf{k_2}, \mathbf{k_3}) \equiv (\mathbf{q}, -\mathbf{j} - \mathbf{q}, \mathbf{j})$. 
    \item In fact, when restricted to the high-$\ell$ regime of $C_{\ell}^{\mathrm{kSZ}^{2}\times\delta_{g}}$, Eq.\eqref{Cl-def} only dictates that $k_3$ must be high (where $k_3\equiv j\approx\ell/\eta$). 
    Consequently, we expect that \textit{two} types of triangle configurations of $\Blos(\mathbf{k_1}, \mathbf{k_2}, \mathbf{k_3})$ contribute significantly to the overall estimator at small scales: (1) Triangles whose all 3 sides are large i.e. high-$k_1,k_2,k_3$, and (2) \textit{Squeezed} triangles whose one side is small and two sides (including $\mathbf{k_3}$) are large. 
\end{itemize}
Both of the above considerations imply that even for astrophysical applications of the projected-fields estimator (e.g. constraining the baryon profile, which largely affects high-$\ell$), it is important to account for all triangle configurations of $\Blos$ (especially the squeezed triangles), and not just those with high-$k_1,k_2,k_3$. In the absence of an assumed limiting scale regime,
\begin{itemize}
    \item A generalized expression must be used for the leading order term of $\Bperp$. Assuming only the Limber approximation, we re-derived the improved expression for $\Pperp$ as given in \cite{Ma2002} (Eq.\,6 therein). After this verification, we analogously derive a rigorous expression for the `usual' term's contribution to $\Bperp$ (denoted by $\Bperp^{\mathrm{usual}}$):
\begin{equation}
  \int \frac{d^{3}k}{\left( 2\pi \right) ^{3}} \left[\sqrt{1 - \mu_{1}^{2}}\sqrt{1 - \mu_{2}^{2}} \right] P_{vv}(k) B_{\delta\delta\delta }( \mathbf{k_{1}}-\mathbf{k}, \mathbf{k_{2}}+\mathbf{k},\mathbf{k_{3}}).
\end{equation}
\end{itemize}
Comparing this to the corresponding expression for D05 in Eq.\eqref{dedeo}, we see that only the geometrical factors (in square brackets) are different, while the rest is the same. Mathematically, our rigorously derived factor $\left[\sqrt{1 - \mu_{1}^{2}}\sqrt{1 - \mu_{2}^{2}} \right]$ is always less than or equal to D05's $[1 - \mu_{1}\mu_{2}]$, with equality occurring only when $\mu_{1} = \mu_{2}$. 
\begin{itemize}
    \item We also include the contributions to $\Bperp$ from the three non-zero `extra' terms. 
\end{itemize}
The $\Bperp^{\mathrm{extra-1}}$ and $\Bperp^{\mathrm{extra-2}}$ terms are symmetric and are effectively the same as $\Bperp^{\mathrm{usual}}$ with an additional scaling of $(k/k_1)$ and $(k/k_2)$, respectively. The $\Bperp^{\mathrm{extra-3}}$ term has a different functional form, and a rough geometric scaling: $( [-k + (k_1$ or $k_2)]/k_3 )$, where either $k_1$ or $k_2$ appear in the numerator depending on how the expression is simplified (both ways are equivalent). Unlike D05's claim, the extra$-3$ term does not have a pure $(k/k_3)$ scaling, since it is associated with the $\delta$ mode (while extra$-1$ and extra$-2$ terms are associated with $p_{\hatbf{n}}$).  

 We refer the reader to Appendix \ref{deriv} for details of our rigorous derivation and a comparison of the four different terms' contributions in an illustrative case. Overall, our improved model is given by $\Blos = (1/2) \left[\Bperp^{\mathrm{usual}}+ \Bperp^{\mathrm{extra-1}} + \Bperp^{\mathrm{extra-2}} + \Bperp^{\mathrm{extra-3}} \right]$, where the expressions for the four terms are given by Eqs. (\ref{usual}, \ref{extra-1}, \ref{extra-2}, and \ref{extra-3}). Since our improved model for $\Blos$ does not make any further approximations, it accurately predicts the theoretical signal $C_{\ell}^{\mathrm{kSZ}^{2}\times\delta_{g}}$ at all scales of interest, which allows us to use the projected-fields estimator for extracting cosmological information. 

\section{Methodology}
\subsection{Survey Specifications} \label{survey}
We now review the specifications of the CMB and LSS experiments that we have assumed in our analysis.  
The projected-fields estimator requires a CMB map that has been `cleaned' or separated from the other contaminating foregrounds in the microwave sky using some multi-frequency component-separation technique. Following the analysis in \cite{Hill2016, F16} for the joint data from \textit{Planck} \cite{Planck2018} and \textit{WMAP} \cite{wmap}, we take the measured CMB temperature map that has been cleaned \cite{LGMCA-Planck} using the ``local-generalized morphological component analysis" (LGMCA) technique \cite{LGMCA}. This map has a Gaussian beam $b(\ell)$ with $\theta_{\mathrm{FWHM}}$ = 5 arcmin, and a white noise power spectrum $N_\ell$ given by
\begin{equation} \label{Pl}
  N_\ell = \Delta_T^2 \, b(\ell)^{-2} = \Delta_T^2 \, \mathrm{exp} \left(\ell(\ell+1)\frac{\theta_{\mathrm{FWHM}}^{2}}{8\,\mathrm{ln}\,2} \right),
\end{equation}
where the ``effective" pixel noise level $\Delta_T = 47 \mu$K-arcmin includes noise from residual foregrounds. In order to obtain the Wiener filter $F(\ell) = C_{\ell}^{\mathrm{kSZ}}/C_{\ell}^{\mathrm{tot}}$ (Eq.~\ref{wiener}), we compute the theoretical kSZ power spectrum using a template\footnote{https://github.com/nbatta/SILC/blob/master/data/ \\
ksz\_template\_battaglia.csv} derived from cosmological hydrodynamical simulations \cite{Battaglia2010}. Here, $C_{\ell}^{\mathrm{tot}}$ = $\left(C_{\ell}^{TT} + C_{\ell}^{\mathrm{kSZ}} + N_\ell\right)$, where the lensed primary CMB temperature power spectrum $C_{\ell}^{TT}$ is calculated for our fiducial model using CAMB \cite{camb1, camb2}. The filter for \textit{Planck} is computed assuming its effective noise (Eq.~\ref{Pl}), with an $\ell_{\mathrm{\max}} = 3000$. 

On the other hand, the upcoming \textit{Simons Observatory} (SO) and the subsequent CMB-S4 experiments will map the CMB at a high resolution of $\theta_{\mathrm{FWHM}}\approx1.4$ arcmin (or $\ell_{\mathrm{\max}} \approx 8000$). We forecast for their cleaned CMB maps that will be obtained using the standard ``Internal Linear Combination" (ILC) component-separation technique \cite{ILC1, ILC2}, which minimizes the total variance. While some earlier forecasts for the kSZ (e.g.\cite{Mueller2014}) and the projected-fields estimator \cite{D05, F16} assumed instrumental noise only, we use the simulated post-ILC noise curves for SO \cite{SO} and CMB-S4 \cite{S4} respectively, which are publicly available online\footnote{https://github.com/simonsobs/so\_noise\_models/tree/master/\newline
LAT\_comp\_sep noise/v3.1.0; standard ILC:deproj-0}${}^{,}$\footnote{https://sns.ias.edu/jch/S4\_190604d\_2LAT\_Tpol\_default\newline\_noisecurves.tgz}. These `ILC' noise power spectra include the detectors' white noise as well as noise due to any residual foregrounds present after component separation, and are thus considerably larger (e.g. see Fig. 4 in \cite{RS_SF}). 

\begin{figure}
\center
{\includegraphics[width=1.05\columnwidth]{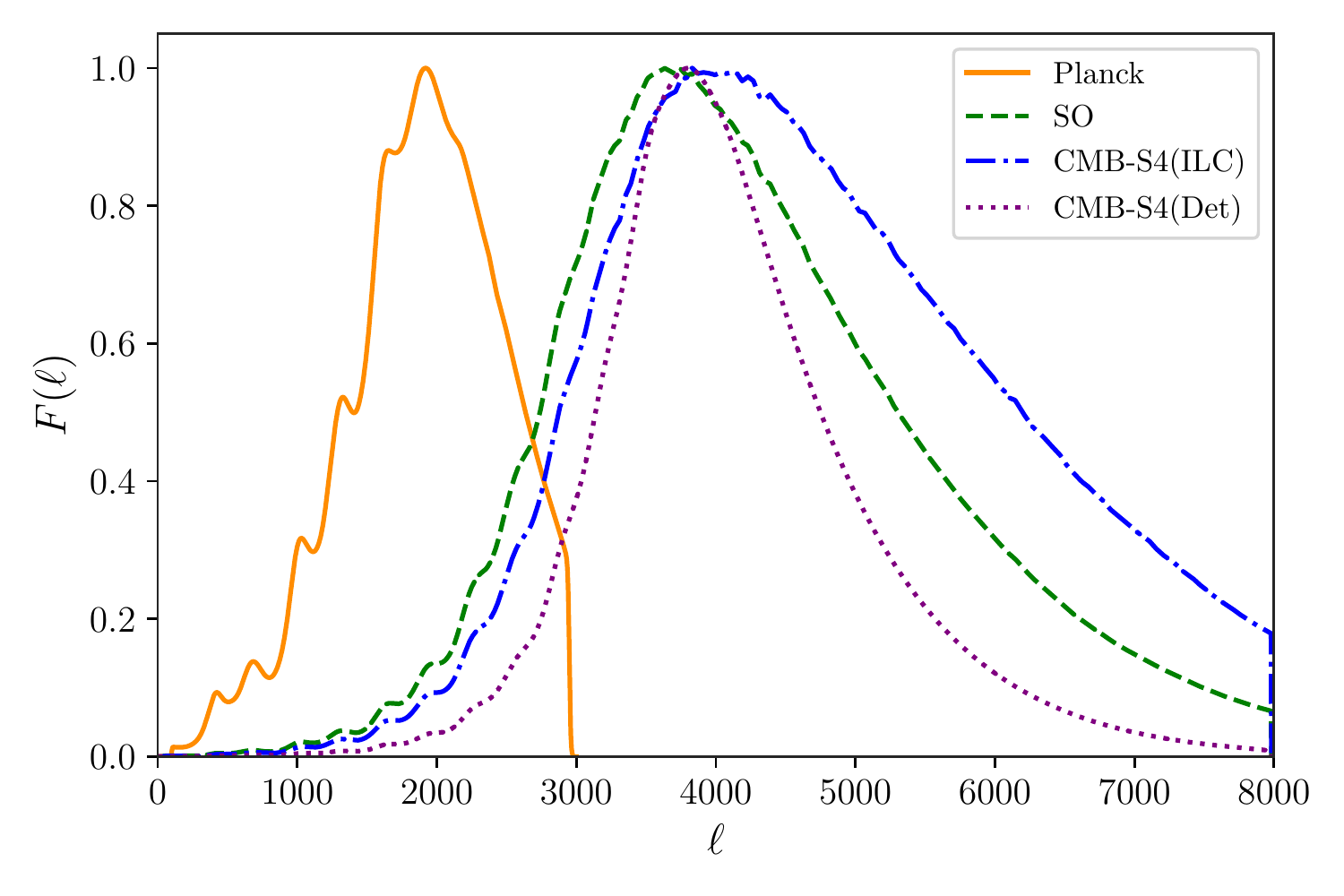} 
}
\caption{Wiener filters (Eq.~\ref{wiener}) $F(\ell)$ for the CMB experiments studied in this work, before multiplying with their corresponding beams as described in Table \ref{tab:cmb_specs}. The normalization here is arbitrarily chosen such that each $F(\ell)$ peaks at 1, and it does not affect any of the results.
}
\label{fig:fl}
\end{figure}

Additionally, we also present forecasts assuming detector-only noise for CMB-S4 (denoted as `CMB-S4 (Det)' henceforth) in Section \ref{impact-fgs} to demonstrate the major impact of residual foregrounds on these forecasts, and to connect them with previous works. Throughout this paper, unless mentioned otherwise, CMB-S4 refers to the CMB-S4 (ILC) case. We construct the standard Wiener filters (Eq.~\ref{wiener}) for SO and CMB-S4 using these noise curves, and the same theoretical $C_{\ell}^{TT}$ and $C_{\ell}^{\mathrm{kSZ}}$ as for \textit{Planck} above. Our filters for these CMB experiments are shown in Figure \ref{fig:fl}, and their specifications are summarized in Table \ref{tab:cmb_specs}.

\begingroup
\begin{table}[!htb]
\begin{center}
\renewcommand{\arraystretch}{1.28} 
\begin{tabular}{|c|c|c|c|c|c|}
\hline
\hline
 CMB&$\theta_{\mathrm{FWHM}}$&Noise $\Delta_T$&Noise&$\ell_{\mathrm{\max}}$&$f_{\mathrm{sky}}$ \\
 experiment&[arcmin]&[$\mu$K-arcmin]&description& & \\ \hline
\textit{Planck}&5&47&effective$^{a}$&3000&0.7\\ 
\hline
SO&1.4&ILC&realistic$^{b}$&8000&0.4\\
CMB-S4 (Det)&3&1&detector-only&8000&0.4\\
CMB-S4 (ILC)&1.4&ILC&realistic$^{b}$&8000&0.4\\
\hline
\hline
\end{tabular}
\end{center}
\caption{Specifications for the forecasted CMB experiments. We assume the $^{a}$effective noise (following \cite{Hill2016, F16}) and $^{b}$post-ILC noise \cite{SO} for \textit{Planck} and SO respectively, which include residual foreground contributions and are thus realistic. We consider two distinct noise models for CMB-S4: post-ILC noise \cite{S4}, and detector-only noise (as assumed for case 3 in \cite{F16}). Unless mentioned otherwise, CMB-S4 refers to the CMB-S4 (ILC) case.}
\label{tab:cmb_specs}
\end{table}

\endgroup

We consider galaxies as a tracer of the LSS (Eq.~\ref{delg}) and include forecasts for two different photometric galaxy surveys. The WISE mission \cite{WISE} imaged the entire sky in the infrared, collecting a catalog of more than 500 million objects. We use the same criteria as \cite{WISE2} to select a sample of galaxies, which were originally studied in \cite{WISE3}. Following the data analysis in \cite{F16, Hill2016}, we similarly assume a mask with $f_{\mathrm{sky}} = 0.447$ and 46.2 million selected galaxies. The redshift distribution of these galaxies peaks at $z \approx 0.3$ and extends up to $z = 1$ \cite{WISE_z} (as plotted in Fig. 10 of \cite{F16}). 

The future VRO (formerly named LSST) \cite{VRO2019} will take deep optical images of about half of the sky, creating catalogs of an unprecedented number of objects. The high-SNR ``gold" sample of LSST (i.e. with $i < 25.3$) will contain around 4 billion galaxies over 20,000 deg$^{2}$, based on empirical estimates from previous surveys. We assume the predicted redshift distribution of these galaxies that extends from $z = 0$ to $z = 3$, with a peak at $z\approx0.7$ \cite{rubin} (Eq.(3.8) therein with $z_{0} \approx 0.3$). We refer the reader to Section 3.7 of \cite{rubin} for further details.

We assume a constant fiducial linear galaxy bias $b_{g} = 1.13$ for WISE galaxies, the best-fit value determined from \textit{Planck} (2015) CMB lensing maps \cite{F16} and the \textit{Planck}$\times$WISE projected-fields measurement \cite{F16, Hill2016}. We allow a redshift-dependent fiducial $b_g$ for VRO galaxies \cite{rubin} given by the model $b_g(z) = 1 + 0.84z$, which has been estimated from simulations in \cite{Weinberg2004}. 

\subsection{Total Covariance Matrix}
Our parameter forecasts depend crucially on the covariance matrix of measurement of the projected-fields estimator. We follow \cite{Dore2004, F16} and assume Gaussianity, so that the total covariance matrix $M_{\ell\ell^'}$ is diagonal and written as:
\begin{equation}\label{cov}
  M_{\ell\ell^'} = \frac{\delta_{\ell\ell^'}}{(2\ell + 1)f_{\mathrm{sky}}} \left[C_{\ell}^{\Bar{T}^2\Bar{T}^2,f} C_{\ell}^{\delta_g \delta_g} + \left(C_{\ell}^{\mathrm{kSZ}^{2}\times\delta_{g}}\right)^{2}\right].
\end{equation}
Here, $f_{\mathrm{sky}}$ is the sky fraction over which the CMB and LSS experiments overlap (with values given in Table \ref{tab:cmb_specs}), and for $C_{\ell}^{\Bar{T}\Bar{T},f} \equiv f^2(\ell)C_{\ell}^{\mathrm{tot}}$,
\begin{equation}
    C_{\ell}^{\Bar{T}^2\Bar{T}^2,f} \approx 2 \int \frac{d^{2}\mathbf{L}}{(2\pi)^2} C_{\ell}^{\Bar{T}\Bar{T},f} C_{|\mathbf{\ell}-\mathbf{L}|}^{\Bar{T}\Bar{T},f}. \nonumber
\end{equation}
The projected galaxy density power spectrum is: \cite{Limber, extlimber}
\begin{equation}
   C_{\ell}^{\delta_g \delta_g} = \int_{0}^{\eta_{\max}} \frac{d\eta}{\eta^2} \, [W^{g}(\eta)]^{2} P_{\delta\delta}\left(k = \frac{\ell}{\eta}, \eta \right) + \frac{1}{\Bar{n}},
\end{equation}
where $\Bar{n}$ is the projected number of galaxies per steradian, appearing in the shot noise term $(1/\Bar{n})$. The first term in brackets in Eq.\eqref{cov} is the contribution due to the actual measurement which includes the instrumental (or ILC) noise of the CMB and LSS experiments. Meanwhile, the second term is the corresponding \textit{cosmic variance} - an inherent statistical uncertainty in the measurement that arises because only one realization among all possible universes described by a certain model can be observed.   

\subsection{Numerical Implementation and the Effect of Massive Neutrinos}\label{num}
In order to calculate the projected-fields estimator $C_{\ell}^{\mathrm{kSZ}^{2}\times\delta_{g}}$ given by Eq.\eqref{Cl-def}, we equivalently compute the LOS integral in the redshift $(z)$ variable, where $\eta(z) \equiv \int_{0}^{z} dz [c/H(z)]$. We discretize the integral and use a coarse bin size $\Delta z = 0.1$ in the window functions $W^{g}(z)$ for the WISE and VRO photometric surveys. The triangle power spectrum $\mathcal{T}$ (Eq.~\ref{triangle}) is calculated independently in each redshift bin, using the corresponding CMB filter $f(\ell)$ and the theoretical hybrid bispectrum $\Blos$. We now summarize our numerical computation of $\Blos$ assuming our \emph{improved} model Eq.\eqref{full}, which involves the computation of the four terms given in Eqs.\,(A.6, A.8-10).

As noted in Section \ref{D05}, we follow \cite{D05} and subsequent works and assume the simulation-based phenomenological substitution suggested by \cite{Zhang2004}, in which we substitute the non-linear matter density field into Eq.\eqref{lin-vel} for every velocity factor appearing in the improved model equations. The resulting non-linear matter power spectrum $P_{\delta\delta}$ is calculated \cite{Takahashi2012} using CAMB \cite{camb1, camb2} for our assumed fiducial model, defined by:
$\{H_{0}, \Omega_{b}h^{2}, \Omega_{c}h^{2}, 10^{9}A_{s}, n_{s}, \tau_{re} \}+\{\Sigma m_{\nu}\}$.
We use the best-fit \textit{Planck}-2018 values (Table 1 of \cite{Planck2018}) for the first six, which are the base parameters in $\Lambda$CDM cosmology: Hubble constant, baryon density, cold dark matter density, amplitude and slope of the primordial spectrum of metric fluctuations, and optical depth to reionization, respectively. 

We also consider an extension to this base model by varying the sum of neutrino masses ($\Sigma m_{\nu}$) in Section \ref{fish-mnu}, which has been confirmed to be non-zero by detections of flavor oscillations in solar and atmospheric neutrinos (e.g. \cite{super-Kamiokande, SNO}). These detections allow for a normal hierarchy of neutrinos with a minimum $\mnus$ of $\sim58$ meV, as well as an inverted hierarchy ($m_3\ll m_1$\,$\simeq$\,$m_2$) with a minimum $\mnus$ of $>100$ meV \cite{mnu_CMBLSS, mnu_whitep}. We assume a normal hierarchy for the neutrinos ($m_1$\,$\simeq$\,$m_2$\,$\simeq$\,0\,$\ll$\,$m_3$) and a fiducial value of 60 meV. Aside from affecting the expansion rate, importantly, massive neutrinos suppress $P_{\delta\delta}$ at scales smaller than their free-streaming scale, defined as $k_{\mathrm{fs}}(z) = 0.018 \,\Omega_{m}^{1/2}(z)[m_{\nu}$/1 eV] h Mpc$^{-1}$.(e.g. \cite{Lesgourgues, mnu_whitep}). 

The $\CTTg$ estimator contains two factors of the peculiar velocity field of the LSS, which is largely dominant at linear and quasi-linear scales. After the phenomenological substitution \cite{Zhang2004} using Eq.\eqref{lin-vel}, the overall signal is proportional to the square of the linear growth rate $f$ of matter perturbations. The presence of massive neutrinos induces a small scale dependence in $f$ that is very well-modeled by the fitting function \cite{Kiakotou2007}: $f(k, z) \approx \mu(k)\Omega_{m}^{\alpha}(z)$ with $\alpha\approx0.55$ for $\Lambda$CDM, and
\begin{equation}
    \mu(k) = 1 - A(k)\Omega_{\Lambda}f_{\nu} + B(k)f_{\nu}^{2} - C(k)f_{\nu}^{3},
\end{equation}
where $f_{\nu} =\Omega_{\nu}/\Omega_{m}$ is the fractional contribution of neutrinos to the total matter density ($\approx$ 0.0045 in fiducial model), and $\Omega_{\Lambda}$ is the dimensionless energy density of dark energy ($\approx 0.6847$ in fiducial model). The coefficients $A(k), B(k), C(k)$ are obtained by interpolating the values given in Table II of \cite{Kiakotou2007}. 

To compute the non-linear matter bispectrum $B_{\delta\delta\delta}$, we follow previous works \cite{F16, Hill2016, Kusiak2021} and use the fitting function from \cite{Gilmarin2012}, which has the form:
\begin{equation}\label{gil}
    B_{\delta\delta\delta}(k_1, k_2, k_3) = \sum_{cyc} 2F_{2}(k_1, k_2, k_3)P_{\delta\delta}(k_1)P_{\delta\delta}(k_2),
\end{equation}
where $\sum_{cyc}$ denotes a cyclic sum over $(k_1, k_2, k_3)$ and $P_{\delta\delta}$ is again the non-linear matter power spectrum. The kernel $F_2$'s scale and redshift dependence is governed by parameters that are fitted using $\Lambda$CDM-only N-body simulations. However, this formula is likely general enough to be applicable at the first-order for certain $\Lambda$CDM-extensions since the kernel is weakly dependent on cosmology \cite{Gilmarin2012}. To compute the matter bispectrum, we use the matter power spectrum in the presence of massive neutrinos in the fitting function above (Eq.~\ref{gil}). At first-order, this is consistent with results from second-order perturbation theory \cite{vlah} and simulations \cite{Ruggeri2017, Coulton} which found that the suppression in the bispectrum's amplitude is almost twice that of the power spectrum. 

Similar to previous works \cite{Hill2016, F16, Kusiak2021}, our numerical computation here implicitly assumes that baryons trace the dark matter distribution (where we have dropped the subscripts on overdensities $\delta$), which is true at large scales ($\sim$Mpc), but not at smaller $\sim$kpc scales (e.g.~\cite{Schaan2021, MayaACT}). Alternatively, the matter bispectrum could be computed within the Halo Model framework \cite{halomodel}, as done recently in \cite{Bolliet2022} assuming the D05 model and using their code \texttt{class_sz}\footnote{\url{https://github.com/borisbolliet/class_sz/releases/tag/v1.0.0}}. We discuss this and other possible future directions in Section V.

The direct numerical evaluation of $\CTTg$ with our improved model is even more time-consuming than the D05 model. While \cite{Bolliet2022} used Fourier transforms to accelerate the approximate computation, the same cannot be done for our model since the integrands for $\Blos$ are not `separable'. In contrast, we speed up by creating emulators for $\Blos$ in each redshift bin using the \texttt{ostrich} library\footnote{\url{https://github.com/dylancromer/ostrich}} \cite{ostrich_cite}, which uses Principal Component Analysis and Gaussian Process Interpolation. The resulting computation of a $\CTTg$ is accurate and relatively efficient, taking $\mathcal{O}$(10 s) on a laptop.  

\subsection{CMB Lensing Contribution}
The projected-fields estimator involves squaring a (cleaned and) filtered CMB temperature map. As a result, the actual measurement gets a contribution due to the weak lensing of the CMB by intervening LSS, in addition to the kSZ signal $\CTTg$ modeled in Section II. The leading-order lensing contribution is given by \cite{F16}:
\begin{align*} 
 \lens \approx \frac{-2\ell C_{\ell}^{\psi\delta_{g}}}{\left(2\pi\right)^2} \int_{0}^{\infty}dL^{'} L^{'2}f(L^{'})C_{L^{'}}^{TT} \int_{0}^{2\pi}d\phi f(|\mathbf{L}^{'}+\mathbf{\ell}|) \, \mathrm{cos}\phi, 
    \numberthis
    \label{lens}
\end{align*}
where $C_{\ell}^{\psi\delta_{g}}$ is the cross-power spectrum between the lensing potential $\psi$ (see \cite{Lewis2006} for a review of CMB lensing) and the tracer field, and $C_{\ell}^{TT}$ is the unlensed primary CMB power spectrum. The lensing convergence, $\kappa(\theta)=-$$\nabla^{2}\psi(\theta)$, so that in Fourier space, $C_{\ell}^{\psi\delta_{g}} = (2/\ell(\ell+1))C_{\ell}^{\kappa\delta_{g}}$. We thus compute $C_{\ell}^{\psi\delta_{g}}$ using the Limber approximation \cite{Limber, extlimber}: 
\begin{equation}
    C_{\ell}^{\kappa\delta_{g}} = \int_{0}^{\infty} \frac{d\eta}{\eta^{2}} \, W^{g}(\eta)W^{\kappa}(\eta) P_{\delta\delta}\left(k = \frac{\ell}{\eta}, \eta \right),
\end{equation}
where $W^{\kappa}(\eta)$ is the CMB lensing kernel (e.g. \cite{sherwin}). 

As seen from Eq.\eqref{lens}, $\lens$ vanishes if the beam is infinitesimally small and no filter is applied \cite{F16}. For the projected-fields measurements with \textit{Planck} data \cite{Hill2016, Kusiak2021}, the lensing contribution is much higher than the kSZ signal \cite{F16}. In contrast, SO and CMB-S4 will have a much smaller beam, and will cover $\ell\sim3000-8000$ where the kSZ effect dominates over the anisotropies due to lensing. Thus, unlike \textit{Planck}, $\CTTg$ will dominate over $\lens$ contribution in future measurements with SO and CMB-S4, as shown in the left panel of Figure \ref{fig:plotratios}.

\section{Results} \label{compare}
\begin{figure*}
\center
{
\includegraphics[width=0.82\textwidth]{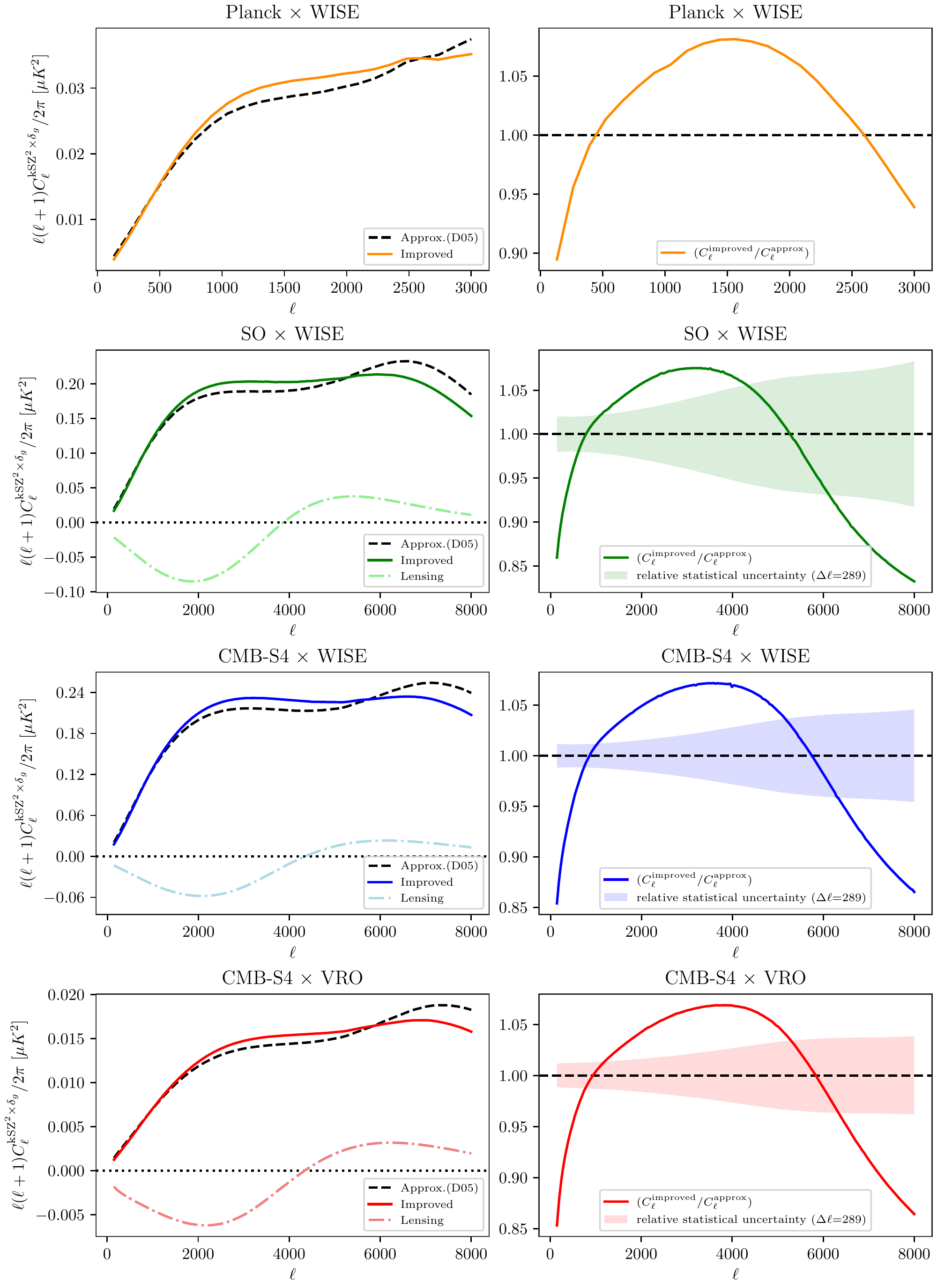}
}
\caption{(left panel): Projected-fields estimator signal as predicted by the approximate `D05' model \cite{D05} ($C_{\ell}^{\mathrm{approx}}$; dashed black lines) and our improved model ($C_{\ell}^{\mathrm{improved}}$; solid colored lines), for (top-to-bottom) \textit{Planck}, SO, and CMB-S4(ILC) data cross-correlated with WISE galaxies, and CMB-S4(ILC) data with VRO galaxies, respectively. The $\lens$ contributions (dot-dashed lines) are shown for all experiments except \textit{Planck}$\times$WISE, for which $\lens$ is much higher than the $\CTTg$ \cite{F16}.\\
(right panel): Ratios of the two models' predicted signals, ($C_{\ell}^{\mathrm{improved}}/C_{\ell}^{\mathrm{approx}}$; solid colored lines). The shaded regions show ratios between the statistical uncertainties (Eq.~\ref{cov}) and the corresponding $C_{\ell}^{\mathrm{approx}}$, with a linear binning of $\Delta\ell = 289$. The relatively large statistical errors for \textit{Planck}$\times$WISE are beyond the axis range shown here \cite{Hill2016}, and are thus omitted.} 
\label{fig:plotratios}
\end{figure*}
In this section, we compare the predictions of our rigorous improved model and the approximate D05 model, for WISE galaxies ($\delta_g$) cross-correlated with \textit{Planck}, SO and CMB-S4 data, and for VRO galaxies cross-correlated with CMB-S4 data. $\CTTg$ assuming our improved model is numerically computed as described in Section \ref{num}. The approximate signal is evaluated similarly, using Eq.\eqref{approx} instead for the hybrid bispectrum, where the matter bispectrum and non-linear $v_{\mathrm{rms}}^2$ are consistently computed with the same assumptions as above. 

The first plot in the left panel of Figure \ref{fig:plotratios} compares the models' predictions for \textit{Planck}$\times$WISE, while the plot on the right shows the ratio between the two. The differences between the predicted signals are up to a $\sim5\%$ level and are thus negligible as compared to the large statistical errors for the \textit{Planck}$\times$WISE measurement \cite{Hill2016}. 
We note that the numerical simulations and the approximate model up to an $\ell_{\mathrm{max}}$ of $\approx3000$ (i.e. \textit{Planck}'s resolution) agree to within $\sim$5-10$\%$ in \cite{F16}. 
Thus, the differences we find from the improved model would not be resolved in previous theory comparisons or impact previous measurements with \textit{Planck} \cite{Hill2016}.
Interestingly, the improved model predicts a higher signal for most $\ell<3000$ and could explain some (but not all) of the excess $\Ak$ found for the green unWISE sample in \cite{Kusiak2021}.

On the other hand, the three lower rows of plots in Figure \ref{fig:plotratios} imply that our improvements in the theoretical model lead to $\sim15\%$ level differences in the predicted signal for the upcoming SO and CMB-S4 experiments, which have enormous forecasted SNRs (see \cite{F16, Bolliet2022}, and our Section \ref{fisher}). The right panel of Figure \ref{fig:plotratios} shows that these differences are significantly larger than the associated statistical uncertainties for these measurements (given by Eq.~\ref{cov}), and hence must be taken into account. 

From Figure \ref{fig:plotratios}, we also see that our improvements in the theoretical model have the largest impact on the predicted $\CTTg$ at both the low-$\ell$ and high-$\ell$ ends, despite the arguments in \cite{Dore2004, D05} justifying the use of the approximate model at small scales. From Section \ref{improv}, the `usual' term in our model for $\Bperp$ is always $\leq \Bperp^{\mathrm{approx}}$, while the other 3 additional terms may or may not contribute comparably depending on the triangle configuration.

Therefore, the ratios of the overall $\CTTg$ shown here can be less than or greater than 1 at different scales, and differ significantly at high-$\ell$s due to the convolution in Eq.\eqref{triangle} and because \textit{squeezed} triangles beyond the high-$k_1,k_2,k_3$ limit contribute significantly to the signal. Moreover, since these differences in the signal are scale-dependent, the corresponding change in the shape of $\CTTg$ would also likely impact parameter inferences. In summary, it is crucial to adopt our improved model in \emph{any} analysis that applies the projected-fields estimator to upcoming high-resolution CMB experiments, to ensure unbiased results. 

\section{Parameter Forecasts} \label{params}

\subsection{Cosmological Dependence}\label{cosmo-dep}
Not only is our improved model necessary to avoid biased inferences from future measurements, but it also enables us to accurately study the cosmological dependence of the $\CTTg$ signal for the first time since it is not restricted to small scales. As noted in Section \ref{num}, this signal depends on the baryon profile, as well as other astrophysics such as Halo Occupation Distribution (HOD) \cite{Bolliet2022} and non-linear bias for galaxies. Here, we focus on the cosmological dependence of the signal and assume a linear galaxy bias $b_g$ for simplicity. The other astrophysical parameters are assumed to be known externally. 

So far, the two \textit{Planck} measurements with this estimator \cite{Hill2016, F16, Kusiak2021} jointly fit for $b_g$ and the amplitude of this kSZ signal, $\Ak$. It is defined as a single free parameter such that $\CTTg= \Ak \CTTg|_{\mathrm{fid}}$; its fiducial value is 1. From Eq.\eqref{Cl-def}, $\Ak\propto(f_{b}f_{\mathrm{free}})^2$, where $f_b$ and $f_{\mathrm{free}}$ are the baryon and free electron fractions at the redshift of the LSS tracers. While we use a redshift-dependent galaxy bias for VRO (Section \ref{survey}), here, we vary the bias amplitude $b_g$ which can be redefined so that its fiducial value matches that of WISE galaxies ($\approx1.13$). The total measured signal is $C_{\ell}^{\mathrm{tot}} = (\CTTg+ \lens)$. Since $\lens\propto b_g$ and $\CTTg\propto b_g\Ak$, there is a degeneracy between $\Ak$ and $b_g$, and some of it is broken by the $\lens$ term \cite{F16}.

We forecast the 1$\sigma$ statistical uncertainty on $\Ak$, denoted by $\Delta\Ak$ $(=\Delta\Ak/\Ak|_{\mathrm{fid}})$, for future measurements with SO$\times$WISE, CMB-S4$\times$WISE, and CMB-S4$\times$VRO, assuming post-ILC CMB noise. We use the Fisher matrix formalism (e.g.\cite{Tegmark1997}), where the Fisher matrix is given by:
\begin{equation}\label{fisher}
    F_{ij} = \sum_{\ell,\ell^{'}} \frac{\partial{C_{\ell}^{\mathrm{tot}}}}{\partial{\theta_{i}}} [M^{-1}]_{\ell\ell^'} \frac{\partial{C_{\ell^{'}}^{\mathrm{tot}}}}{\partial{\theta_{j}}}, 
\end{equation}
and the covariance matrix of the measurement $M_{\ell\ell^'}$ is given by Eq.\eqref{cov}. For this Fisher matrix, the marginalized $1\sigma$ uncertainty on each parameter $\theta_{i}$ is given by $\sqrt{[F^{-1}]_{ii}}$. 

\begin{figure}
\center
{
\includegraphics[width=0.98\columnwidth]{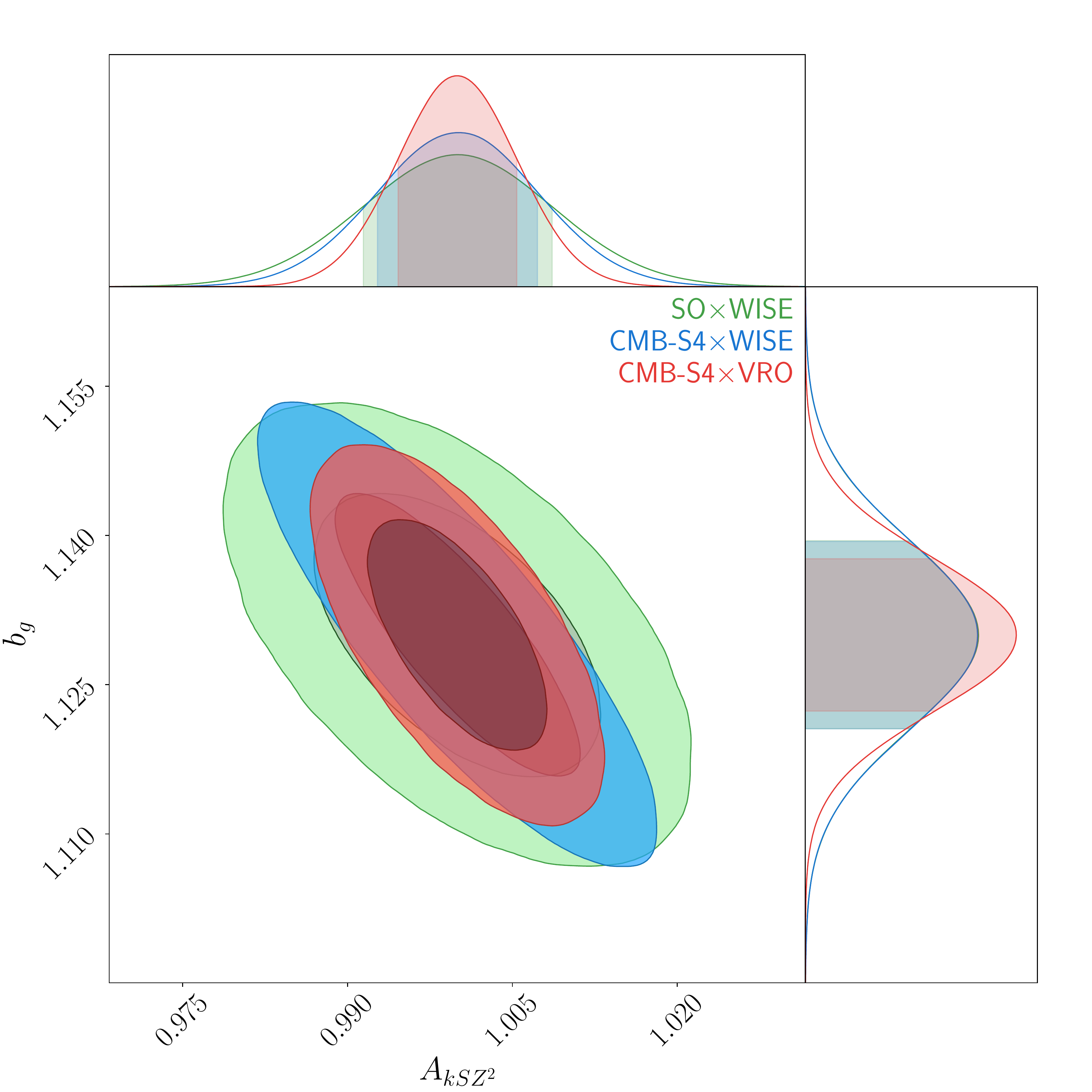} 
}
\caption{Contours show 68$\%$ confidence levels ($1\sigma$) for $\{\Ak, b_g\}$ for forecasted projected-fields kSZ measurements with SO$\times$WISE (green), CMB-S4$\times$WISE (blue), and CMB-S4$\times$VRO (red). Here, we assume realistic post-ILC noise for SO and CMB-S4, a 1$\%$ Gaussian prior on $b_g$, and \textit{fixed} cosmological parameters.} 
\label{fig:SNR_contours}
\end{figure}

As in previous measurements, we first assume that the cosmology is \textit{fixed}, so the parameter set is $\pmb{\theta}_{\mathrm{base}} = \{\Ak, b_g\}$. Here, partial derivatives in Eq.\eqref{fisher} are calculated directly due to the linear dependence. We also set a 1$\%$ Gaussian prior on $b_g$, since it can be determined by galaxy auto-correlation ($C_{\ell}^{\delta_g \delta_g}$) and galaxy-lensing cross-correlation ($C_{\ell}^{\kappa \delta_g}$) external measurements (e.g. \cite{F16,unwise}). Figure \ref{fig:SNR_contours} shows the resulting 68$\%$ C.L. contours. Marginalized 1$\sigma$ uncertainties on $\Ak$ are at the sub-percent level for all 3 survey combinations (Table \ref{table:SNRs}: top row), corresponding to an SNR $(\equiv \Ak/\Delta\Ak)$ of about 113, 127, and 182 for SO$\times$WISE, CMB-S4$\times$WISE, and CMB-S4$\times$VRO, respectively. 

\begingroup
\begin{table*}
\setlength{\tabcolsep}{3.5pt}
\renewcommand{\arraystretch}{1.28} 
        \begin{tabular}{|c|c|c|c|}
        \hline
        \hline
 &\multicolumn{3}{|c|}{$\%\Delta\Ak$: 1$\sigma$ $\%$ uncertainty on $\Ak$}\\ \cline{2-4}  
 &SO $\times$ WISE &CMB-S4(ILC)$\times$ WISE&CMB-S4(ILC)$\times$ VRO\\   \hline
 Fixed Cosmology &0.9 &0.8 &0.6 \\ \hline
 Varying $\Lambda$CDM + \textit{Planck} prior&7.1 &6.7 &6.9 \\ 
 Varying $\Lambda$CDM + (low-$\ell$ \textit{Planck}+high-$\ell$ CMB-S4) prior&5.9 &5.6 &5.9 \\ 
 Varying $\Lambda$CDM + (low-$\ell$ LiteBIRD+high-$\ell$ CMB-S4) prior&3.7 &3.6 &3.6 \\ 
        \hline
        \hline
\end{tabular} 
\caption{Marginalized 1$\sigma$ \% uncertainty on $\Ak$ for the three survey combinations considered, when cosmological $\Lambda$CDM parameters are: fixed at their best-fit \textit{Planck}-2018 values \cite{Planck2018} (top row), allowed to vary around their fiducial values with a known \textit{Planck} prior (second row), with a forecasted (low-$\ell$ \textit{Planck} + high-$\ell$ CMB-S4) prior (third row) and with a forecasted (low-$\ell$ LiteBIRD + high-$\ell$ CMB-S4) prior (bottom row) \cite{FishLSS}. In all cases here, we assume an external 1$\%$ Gaussian prior on $b_g$, realistic post-ILC noise, and $\mnus$ fixed at 60 meV.
}
\label{table:SNRs}
\end{table*}
\endgroup
Next, we consider the cosmological dependence of this kSZ signal, and allow the $\Lambda$CDM parameters to vary about their fiducial \textit{Planck}-2018 values. Thus, we compute kSZ Fisher matrices for the parameter set:
\begin{equation}\label{thetavec}
\pmb{\theta}_{\mathrm{min}} = \{H_{0}, \Omega_{b}h^{2}, \Omega_{c}h^{2}, 10^{9}A_{s}, n_{s}\}+\{\Ak, b_g\},
\end{equation}
where we numerically compute partial derivatives with respect to $\Lambda$CDM parameters using the same step-sizes as \cite{Madhavacheril2017} and \cite{cromer_table} (Table I therein), and check for numerical stability. We apply an external prior based on current \textit{Planck} measurements of the CMB (TT,TE,EE+lowE; no CMB lensing) \cite{Planck2018}, obtained using the code \texttt{FishLSS}\footnote{https://github.com/NoahSailer/FishLSS/tree/master/FishLSS/} \cite{FishLSS}, which we marginalize over $\tau_{re}$. 

From Table \ref{table:SNRs}, with this \textit{Planck} prior, the marginalized uncertainties on $\Ak$ are around $\sim7\%$ for all 3 experiments. Compared to the fixed cosmology case, this pronounced increase in $\Delta\Ak$ arises due to the high sensitivity of $\CTTg$ to cosmology, particularly $\Omega_{c}h^2$ and $A_s$, as we explain in detail in Appendix \ref{cosmodep}. To understand this intuitively, note that Eqs.\eqref{Cl-def}, \eqref{lin-vel}, and \eqref{gil} imply a rough scaling for the signal,
\begin{equation} \label{eq-scaling}
    \CTTg\propto (faH/\eta)^{2}\,P_{\delta\delta}^{3}. 
\end{equation}
Even if the bracketed growth factor above is neglected, $P_{\delta\delta} \propto \sigma_{8}^{2-3}$, where $\sigma_{8}$ is the power spectrum normalization. Hence, $\CTTg\propto \sigma_{8}^{6-7}$, as suggested roughly in \cite{Dore2004}. Moreover, $\sigma_{8}$ is a derived parameter mainly dependent on $\Omega_{m}$ (and thus $\Omega_{c}h^2$) and $A_s$. Current \textit{Planck} constraints on $\sigma_{8}$ are $\sim0.9\%$, so this scaling contributes around $5.7\%$ to $\Delta\Ak$.

We note that the strong cosmological dependence of the signal highlighted above does not change the \textit{detection significance} of the kSZ signal, which will be very high, but it will limit the interpretation of the measured amplitude in terms of astrophysical quantities (such as baryon fraction). 

With a basic CMB prior combining information from low-$\ell$ ($\ell<30$) \textit{Planck} data and future high-$\ell$ ($30< \ell<5000$) CMB-S4 data \cite{FishLSS}, the forecasted uncertainties on $\Delta\Ak$ improve to $\sim6\%$ (Table \ref{table:SNRs}). The upcoming LiteBIRD satellite will map the CMB at $\ell<200$ with high sensitivity, allowing a cosmic variance-limited measurement of $\tau_{re}$ \cite{litebird}. Applying a forecasted (low-$\ell$ LiteBIRD + high-$\ell$ CMB-S4) prior on $\Lambda$CDM, $\Delta\Ak$ improve further to $\sim3.5\%$, since $A_s$ and $\tau_{re}$ are highly degenerate in primary CMB data, and $\CTTg$ is extremely sensitive to $A_s$ (Appendix \ref{cosmodep}).

\subsection{Fisher forecasts including Neutrino Masses} \label{fish-mnu}
\begin{figure*}
\center
{
\includegraphics[width=0.96\textwidth]{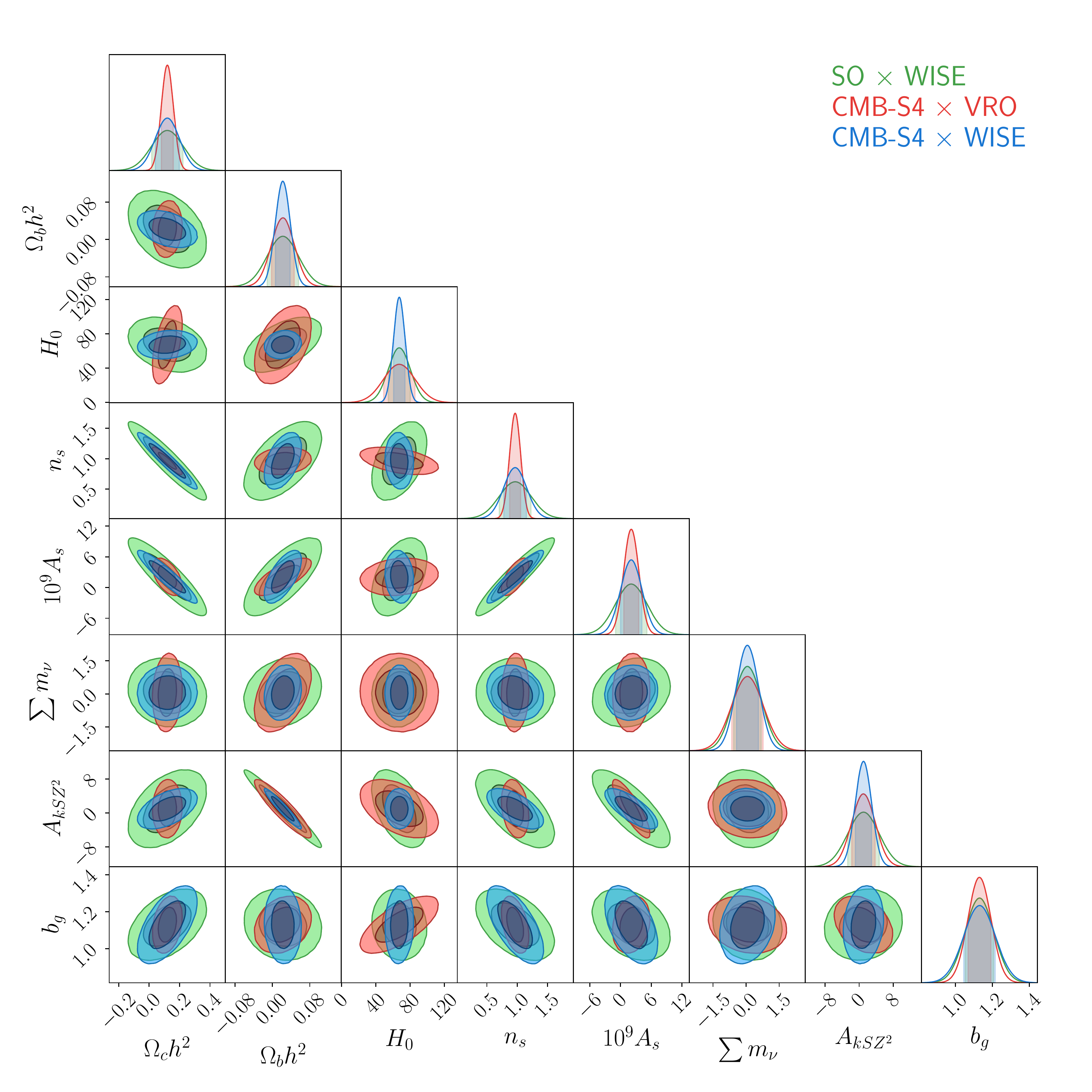} 
}
\caption{Fisher forecasts for marginalized errors on $\Lambda$CDM parameters, $\mnus$ (in eV here), and $\{\Ak, b_g\}$ from kSZ-only projected-fields constraints, without assuming \emph{any} priors, and with realistic post-ILC noise for SO and CMB-S4. Contours show the 2D marginalized 68$\%$ confidence ellipses for SO$\times$WISE (green), CMB-S4$\times$WISE (blue), and CMB-S4$\times$VRO (red).}
\label{fig:Fisher}
\end{figure*}
Equipped with our improved model, we now illustrate the potential sensitivity of this kSZ estimator to extensions of the Standard Model, by studying the dependence of $\CTTg$ on the sum of neutrino masses. Here, we compute kSZ Fisher matrices for the parameters $\pmb{\theta}_{\mathrm{ext}} = \pmb{\theta}_{\mathrm{min}}+\{\mnus\}$. We note that there is no marginalization over higher-order bias parameters or astrophysical parameters related to the halo profile performed here. Therefore, when interpreting these forecasts they should only be taken as representative of the parameter sensitivity of the kSZ$^2$ observable, and would only be attainable if the bias and astrophysical parameters were known externally (for example from a joint analysis with other probes).

At first, we do not apply any external prior. Contours in Figure \ref{fig:Fisher} show the resulting 2D marginalized 1$\sigma$ uncertainties for each pair of parameters, thus informing about their degeneracies for $\CTTg$. CMB-S4$\times$WISE gives tighter constraints than SO$\times$WISE on all parameters except for $b_g$, due to CMB-S4's lower noise. The relative scale-dependence of $\lens$ and $\CTTg$ together determine the error on $b_g$; the relative amplitude of $\lens$ is higher for SO$\times$WISE (Figure \ref{fig:plotratios}).

In all the forecasts presented here, we do not account for the cosmological dependence of $\lens$, but only its linear dependence on the galaxy bias. This is a reasonable assumption since a template for $\lens$ can be obtained separately by cross-correlating the galaxy sample with a CMB lensing map. However, as a check, we also performed an alternative Fisher analysis including the dependence of $\lens$ on all $\Lambda$CDM+$\{\mnus\}$ parameters. In that scenario, forecasted errors on all parameters shift within a margin of $\sim10\%$ for all 3 experiments; we omit the exact values here for brevity. 

\begingroup
\begin{table*}
\centering
\setlength{\tabcolsep}{3.5pt}
\renewcommand{\arraystretch}{1.28} 
         \begin{tabular}{|c|c|c|c|}
         \hline
         \hline
 &\multicolumn{3}{|c|}{kSZ + \textit{Planck} + 1\% $b_g$ prior}\\ \cline{2-4}  
 &SO $\times$ WISE &CMB-S4(ILC)$\times$ WISE&CMB-S4(ILC)$\times$ VRO \\   \hline
     $\sigma(\Sigma m_{\nu})$ [meV]& 168 &100 &254 \\ \hline\hline 
     $\%\Delta\Ak$ &10.9 &8.1 &20.7\\ 
         \hline
         \hline
 \end{tabular} 
 \caption{Forecasts for marginalized 1$\sigma$ errors on $\mnus$ (top row) and on $\Ak$ (written as a \%; bottom row) from the $\CTTg$ kSZ signal jointly constraining the $\pmb{\theta}_{\mathrm{ext}}$ parameters, after applying our \textit{Planck} prior \cite{FishLSS} on $\Lambda$CDM parameters only (given in Table \ref{tab:params-det}). We also set an external 1$\%$ Gaussian prior on $b_g$, and assume realistic post-ILC CMB noise. 1$\sigma$ errors on $\Lambda$CDM parameters improve only marginally over the \textit{Planck} prior, and are thus omitted here.}
 \label{table:paramserror-ksz+pl}
 \end{table*}
 \endgroup
From Figure \ref{fig:Fisher}, we show that the forecasted errors on $\Omega_{c}h^{2},10^{9}A_{s},n_{s}$, and $b_g$ are smaller for CMB-S4$\times$VRO than CMB-S4$\times$WISE, since $\CTTg$ with the former combination is more sensitive to those parameters than the latter. However, CMB-S4$\times$WISE constrains the other four parameters: $\Omega_{b}h^{2},H_{0},\mnus$, and $\Ak$ more tightly. Note that when cosmology is assumed to be fixed (in Section \ref{cosmo-dep} above), the SNR for CMB-S4$\times$VRO is higher than CMB-S4$\times$WISE, and the corresponding $\Delta\Ak$ is smaller with VRO (Figure \ref{fig:SNR_contours}). 

Once cosmological parameters are freed up, i.e. $\Lambda$CDM in Table \ref{table:SNRs} and $\Lambda$CDM+$\{\mnus\}$ in Table \ref{table:paramserror-ksz+pl}, the uncertainty on $\Ak$ is higher with VRO than with WISE, even though VRO is a deeper survey. \cite{Dore2004} shows that differential contribution to $\CTTg$ comes largely from lower-redshift bins peaking around $z\sim0.5$, which reasonably matches the $z$-range of WISE. For deeper surveys such as VRO that have a long tail at the high-$z$ end of their distribution, using an optimal redshift-weighting in $W_{g}$ for $\CTTg$ may possibly improve their SNR \cite{F16} and their cosmological constraints. Alternatively, a low-$z$ subsample of the galaxy survey may be chosen to improve such constraints; we leave this direction for future work.  

We now apply the same \textit{Planck} prior \cite{FishLSS} as the previous subsection on only $\Lambda$CDM parameters, marginalizing over $\tau_{re}$, and include a 1\% Gaussian prior on $b_g$. Table \ref{table:paramserror-ksz+pl} shows the resulting 1$\sigma$ \% uncertainties on $\Delta\Ak$, which are higher than those in Table \ref{table:SNRs} due to the additional marginalization over $\mnus$. Constraints on $\Lambda$CDM parameters improve only slightly over the \textit{Planck} prior. Our results are robust to the choice of the assumed neutrino hierarchy. 

We do not include \emph{any} external prior on $\mnus$ here, and forecast a 1$\sigma$ constraint of $\sim168$ meV and 100 meV on $\mnus$ with the projected-fields kSZ estimator for SO$\times$WISE and CMB-S4$\times$WISE respectively. In comparison, \cite{Mueller2014} forecasted a $\sigma(\mnus)$ of $\sim220$ meV and 96 meV for a SO-like and CMB-S4-like experiment respectively, by combining with spectroscopic galaxy surveys using the pairwise-kSZ estimator, where they included a forecasted \textit{Planck} prior on $\Lambda$CDM as well as $\mnus$. We further discuss our $\sigma(\mnus)$ constraints in the context of other probes in the next subsection.

In our initial analysis here, we demonstrate the potential of this estimator as a probe of cosmology and of the sum of neutrino masses, through their effect on the growth of structure. A more accurate analysis would include a non-linear galaxy bias that also models the induced scale-dependence due to $\mnus$ (e.g. \cite{dvorkin, bis_mnu}), instead of the simple linear galaxy bias $b_g$ assumed here. As discussed in Section \ref{num}, the matter bispectrum too could be numerically computed with a more accurate model, that also possibly accounts for the effect of neutrinos more directly (e.g. \cite{bis_mnu, Ruggeri2017}). 

\subsection{Impact of Residual Foregrounds}\label{impact-fgs}
\begin{table}[]
\begin{center}
\renewcommand{\arraystretch}{1.28} 
\setlength{\tabcolsep}{4.2pt}
\begin{tabular}{|c|c|c|c|}
  \hline
  \hline
 &\multicolumn{3}{|c|}{1$\sigma$ errors on parameters} \\ \cline{2-4}  
 Parameter&Planck&\multicolumn{2}{|c|}{CMB-S4(Det)$\times$WISE} \\ \cline{3-4}  
 &prior&kSZ-only&kSZ + \textit{Planck}\\\hline
$H_0$ [km/s/Mpc]&0.57&3.20 &0.34\\
$\Omega_{b}h^{2}$&0.00013&0.00430 &0.00010\\
$\Omega_{c}h^{2}$&0.0013&0.0131 &0.0008\\
$10^{9}A_s$&0.031&0.366&0.023\\     
$n_s$&0.003&0.030 &0.003\\ \hline
$\Sigma m_{\nu}$ [meV]&-&136 &38 \\ \hline
$\Ak$&-&0.411 &0.025\\
$b_g$&-&0.011 &0.007\\
\hline
\hline
\end{tabular}
\end{center}
\caption{Forecasts for marginalized 1$\sigma$ errors on the $\pmb{\theta}_{\mathrm{ext}}$ parameters assuming detector-only noise and the absence of residual foregrounds in CMB-S4 maps. The third column shows kSZ-only 1$\sigma$ constraints with the $\CTTg$ signal from CMB-S4(Det)$\times$WISE. We include a \textit{Planck} prior \cite{FishLSS} as described in Section \ref{cosmo-dep} on $\Lambda$CDM only (second column), and the resulting kSZ+\textit{Planck} constraints are given (last column).}
\label{tab:params-det}
\end{table}

In the absence of accurate post-ILC noise models for SO and CMB-S4, earlier Fisher forecasts for the kSZ (e.g.~\cite{Dore2004, D05, F16, Mueller2014}) often assumed detector-only white noise. In order to connect with these previous works, in this subsection, we forecast for a CMB-S4(Det)$\times$WISE measurement assuming this noise model for CMB-S4 (Table \ref{tab:cmb_specs}). 

Similar to Section \ref{cosmo-dep}, we first keep the cosmology fixed and forecast an SNR $(\equiv \Ak/\Delta\Ak)$ of $\sim345$, consistent with the \cite{F16} forecast (Table II therein; case 3). Thus, comparing with the CMB-S4(ILC)$\times$WISE case, we see that the presence of residual foregrounds in post-ILC CMB maps diminishes the significance of this detection by a factor of $\sim3$. 

We then vary the $\Lambda$CDM+$\{\mnus\}$ parameters as well, and include the same \textit{Planck} prior \cite{FishLSS} as described in Section \ref{cosmo-dep} on $\Lambda$CDM. The resulting kSZ+\textit{Planck} marginalized errors improve upon the \textit{Planck} errors significantly, unlike the CMB-S4(ILC)$\times$WISE case. This again shows the enormous impact of residual contaminants in post-ILC CMB maps. In the idealistic case of detector-only noise, we forecast a $\sigma(\mnus)$ of 38 meV for CMB-S4(Det)$\times$WISE, compared to 100 meV for CMB-S4(ILC)$\times$WISE. However, the projected-fields $\CTTg$ estimator does need cleaned CMB maps. Multi-frequency techniques such as ILC remove the bulk of foregrounds, but leave some residual contamination. Thus, the CMB-S4(ILC)$\times$WISE forecasts are realistic for this future measurement. 

Although we do not include \emph{any} prior on $\mnus$, our realistic $\sigma(\mnus)$ forecast is tighter than the 95\% C.L. bound of 240 meV from \textit{Planck}'s extended (TT,TE,EE+lowE)+lensing analysis. Assuming detector-only noise, a TT,TE,EE+lowE)+lensing analysis for CMB-S4 optimistically forecasts a $\sigma(\mnus)=73$ meV (e.g.~\cite{mishra2018}), which is again insufficient alone for a significant $\mnus$ detection. Combining it with upcoming BAO measurements from DESI \cite{DESI} or with clustering and shear data from VRO, would allow a 2-3$\sigma$ significant detection of the normal hierarchy (e.g.~\cite{mishra2018,mnu_CMBLSS,mnu_whitep}). The $\CTTg$ kSZ signal is a complementary probe of $\mnus$ and can possibly be combined with these LSS probes to improve upon their constraints.

Moreover, note that this estimator only requires the removal of foregrounds that are correlated with the considered LSS tracers \cite{F16}, so for WISE galaxies, higher-$z$ foregrounds such as the CIB (whose bulk emission comes from $z\gtrsim1$ \cite{cib}) may be neglected. Also, \cite{Kusiak2021} used a different `$\alpha$-cleaning' method which maximizes their SNR. Therefore, such alternative cleaning techniques that are more suited for the $\CTTg$ estimator may give parameter constraints that are intermediate between the two CMB-S4(Det) and CMB-S4(ILC) cases that we have considered here. 

\subsection{Future Directions}\label{discuss} 
Apart from the future directions described earlier in this section, there are a few key ways to build upon our analysis here. As noted in Section \ref{num}, baryon and dark matter density profiles are found to differ at small scales ($\lesssim$ kpc) due to feedback and energy injection, and neglecting this can introduce systematics in the future high-resolution regime. Recently, \cite{Bolliet2022} forecasted constraints on the baryon profile using $\CTTg$, although they are limited in the sense that they assume the approximate D05 model and a fixed cosmology. Incorporating our improved model within the Halo Model approach of \cite{Bolliet2022} that uses a more realistic baryon profile (e.g.~\cite{Battaglia2016}) would thus enable improved forecasts. 
The baryon profile can also be determined externally, for example, with kSZ measurements using velocity-weighted stacking \cite{Schaan2021}.

Another systematic is the galaxy bias that becomes non-linear at small scales, while $\Lambda$CDM extensions such as massive neutrinos induce an additional scale-dependence in it \cite{dvorkin, bis_mnu}. Thus, a non-linear galaxy bias should be included in future analyses for further accuracy. Alternatively, this effect could be consistently accounted for via HOD parameters within the Halo Model framework. 

\section{Conclusions}\label{conclusions} 
In this work, we have considered the projected-fields kSZ estimator which has the distinct advantage of not requiring accurate spectroscopic redshifts. The original formulation of this estimator \cite{Dore2004, D05} has an approximate model (D05) only applicable at small scales, which has been used to estimate the baryon fraction from the signal amplitude $\Ak$ with \textit{Planck} and WISE data \cite{Hill2016, Kusiak2021}. Here, we have presented a first rigorous derivation of an \emph{improved} theoretical model for this estimator, that is thus accurate even at larger scales where cosmological effects are significant. While the differences between the predicted signal from our model and the D05 model are negligible for previous \textit{Planck} detections, they are significant at all scales ($\sim$10-15$\%$) for upcoming measurements with SO and CMB-S4 that are forecasted to have high SNR ($>100$). Given the scale-dependence of these differences, it is crucial to adopt our improved model in \emph{any} future analysis with this estimator to ensure unbiased results.

Our improved model enables us to accurately study the cosmological dependence of the $\CTTg$ signal. While the marginalized uncertainty on $\Ak$ assuming fixed cosmology is at the sub-percent level for future measurements, it increases to $\sim7\%$ when $\Lambda$CDM parameters are freed up with a \textit{Planck} prior, limiting the signal's interpretation. The detection significance remains high, and $\Delta\Ak$ will improve with reduced uncertainty on $\Lambda$CDM parameters (especially $\tau_{re}$) from CMB-S4 and LiteBIRD. As an illustration of the potential sensitivity of this signal to $\Lambda$CDM extensions, our initial analysis forecasts a $\sigma(\mnus)$ of $\sim168$ meV and 100 meV for SO and CMB-S4 (with WISE, and assuming realistic CMB noise), similar to corresponding constraints from pairwise-kSZ \cite{Mueller2014}. This reaffirms the kSZ as a complementary probe of neutrino masses among other CMB and LSS probes. 

While it is important to marginalize over cosmology in any analysis with this estimator, the numerical computation using the improved model will need to be sped up for a full MCMC analysis in the future. Building upon our current method of emulators, a faster implementation may be possible within the \texttt{cosmopower} emulation framework \cite{cosmopower, cosmopowerboris} in the future. Our work is essential for future extended analyses jointly fitting both baryonic astrophysics and cosmology with this kSZ estimator. This could be achieved by incorporating our improved model and its cosmological dependence within the Halo Model framework of \cite{Bolliet2022}. We note that the formalism of this estimator, further developed here, is directly applicable for analyses with any tracers of the underlying matter density, such as quasars, weak lensing convergence (e.g.~\cite{Bolliet2022}), or 21-cm fluctuations probing patchy reionization (e.g.~\cite{Ma2018, LaPlante2022}). 

\begin{acknowledgments}
We thank Boris Bolliet, J. Colin Hill, Aleksandra Kusiak, Mathew Madhavacheril, and Anirban Roy for their helpful feedback and suggestions on this work. We thank Emily Moser and Dylan Cromer for their help in speeding up our computation using the \texttt{ostrich} emulator library. R.P and N.B acknowledge the support from NSF grant AST-1910021 and NASA grant 17-ATP17-0060 for this work and N.B acknowledges additional support from NASA grants 21-ADAP21-0114 and 21-ATP21-0129. S.F. is supported by the Physics Division of Lawrence Berkeley National Laboratory.
\end{acknowledgments}

\appendix
\section{Derivation of our improved Model} \label{deriv}
In Section II, we describe the framework for calculating the hybrid bispectrum $\Blos$, and summarize our improved model and its features. Here, we outline its derivation.   
\subsection{A useful lemma}
Following \cite{Ma2002}, each of the four non-zero terms in Table \ref{tab:wicks} contributing to $\Blos$ (= $\Bperp$/2) can be written in index notation as:\,$\Bperp(\mathbf{k_{1}}, \mathbf{k_{2}}, \mathbf{k_{3}}) = \hatbf{p}_{\perp;1}^{i} \hatbf{p}_{\perp;2}^{j} B_{pp\delta}^{ij}(\mathbf{k_{1}}, \mathbf{k_{2}}, \mathbf{k_{3}})$,
where $\mathbf{p}_{\perp;\alpha}$ is the curl or transverse component of $\mathbf{p}(\mathbf{k}_{\alpha})$ for $\alpha = 1, 2$. So, by definition, $\mathbf{p}_{\perp;\alpha}$ must be perpendicular to the wavevector $\mathbf{k}_{\alpha}$. As noted earlier (and in \cite{Ma2002, Dore2004, D05}), in the Limber approximation, only those modes of momenta with wavevectors $\mathbf{k}_{\alpha}$ perpendicular to the LOS direction $\hatbf{n}$ contribute to the integrated kSZ effect. Moreover, as noted previously in \cite{D05} under the Limber approximation, the orthogonality of $\mathbf{p}_{\perp;\alpha}$ to the LOS gives us the effective relation: $\hatbf{p}_{\perp;\alpha} \approx \hatbf{k}_{\alpha} \times \hatbf{n}$, upto a sign difference. 

Suppose we define the LOS as the z-direction in 3D: $\hatbf{n} \equiv  \hatbf{z}$. For any unit wavevector $\hatbf{m}$, consider the dot product $\hatbf{p}_{\perp;\alpha}^{i}\hatbf{m}^{i} = \hatbf{p}_{\perp;\alpha} \cdot \hatbf{m}$: 
\begin{dmath}
 \hatbf{p}_{\perp;\alpha} \cdot \hatbf{m} 
 = (\hatbf{k}_{\alpha} \times \hatbf{z}) \cdot \hatbf{m} \hspace{1.5cm} \text{[i.e. scalar triple product]} 
 = \hatbf{m}^{x}\hatbf{k}_{\alpha}^{y} - \hatbf{m}^{y}\hatbf{k}_{\alpha}^{x} 
 \approx |\hatbf{m} \times \hatbf{k}_{\alpha}| \hspace{2cm} \text{[Limber Approximation]}
 = \pm \sin(\theta_{\alpha}) \hspace{1.8cm} \text{$\theta_{\alpha} \equiv$ angle between $\hatbf{k}_{\alpha}$ and $\hatbf{m}$} 
= \sqrt{1 - \mu_{\alpha}^{2}} \hspace{1.8cm} {\mu_{\alpha} \equiv \hatbf{k}_{\alpha} \cdot \hatbf{m}}, 
\label{ofac}
\end{dmath}
upto a sign difference. As a verification, we first apply this lemma to derive the general expression for the power spectrum $\Pperp$, which matches exactly with the one given in \cite{Ma2002} (see Eq.\,(6) therein; note that the scaling of the (negligible) connected four-point term also has an analogous form). We then use this relation to calculate the geometric scalings of $\Bperp$ below. 

\subsection{The usual term}
For the leading order `usual' term of $\Bperp$ given in the first row of Table \ref{tab:wicks}, we have
\begin{dmath}
 \langle vv\rangle \langle \delta \delta \delta \rangle = \hatbf{p}_{\perp 1}^{i} \hatbf{p}_{\perp 2}^{j} \int \frac{d^{3}k}{\left( 2\pi \right) ^{3}}\int \frac{d^{3}k^{'}}{\left( 2\pi \right) ^{3}} \\
     \langle v^{i}(\mathbf{k}) v^{j}(\mathbf{k'}) \rangle 
     \langle  \delta( \mathbf{k_{1}}-\mathbf{k}) \delta( \mathbf{k_{2}}-\mathbf{k{'}}) \delta(\mathbf{k_{3}}) \rangle  
\end{dmath}
We substitute the corresponding power spectrum and bispectrum in place of the cross-correlations, similar to Eq.\eqref{Fourier}. Using the sampling property of the Dirac delta function, we eliminate $\mathbf{k'}$ and substitute it with 
(-$\mathbf{k}$):
\begin{align}
    & \langle vv\rangle \langle \delta \delta \delta \rangle \nonumber \\
    & = \left( 2\pi \right) ^{3}\delta _{D}( \mathbf{k_{1}}+\mathbf{k_{2}}+\mathbf{k_{3}}) \times \nonumber \\
    & \int \frac{d^{3}k}{\left( 2\pi \right) ^{3}} [\hatbf{p}_{\perp 1}^{i} \hatbf{p}_{\perp 2}^{j} \hatbf{k}^{i} \widehat{(-\mathbf{k})}^{j}] P_{vv}(k) B_{\delta\delta\delta }( \mathbf{k_{1}}-\mathbf{k}, \mathbf{k_{2}}+\mathbf{k},\mathbf{k_{3}})  
\end{align}
By canceling the Dirac delta enforcing closed triangles on both sides, we get
\begin{align}
    & \Bperp^{\mathrm{usual}}(\mathbf{k_{1}}, \mathbf{k_{2}}, \mathbf{k_{3}}) \nonumber \\
    & = \int \frac{d^{3}k}{\left( 2\pi \right) ^{3}} [\hatbf{p}_{\perp 1}^{i} \hatbf{p}_{\perp 2}^{j} \hatbf{k}^{i} \widehat{(-\mathbf{k})}^{j}] P_{vv}(k) B_{\delta\delta\delta }( \mathbf{k_{1}}-\mathbf{k}, \mathbf{k_{2}}+\mathbf{k},\mathbf{k_{3}})
\end{align}
Consider the orientation factor within square brackets in the equation above. Let the angle between $\hatbf{k}$ and $\hatbf{k}_{\alpha}$ be $\theta_{\alpha}$ for $\alpha$ = 1, 2. Then, by the lemma in Eq.\eqref{ofac}, the geometric scaling of the usual term is
\begin{dmath}
(\hatbf{p}_{\perp;1}^{i} \hatbf{k}^{i}) (\hatbf{p}_{\perp;2}^{j} \widehat{(-k)}^{j}) = (\hatbf{p}_{\perp;1} \cdot \hatbf{k}) (\hatbf{p}_{\perp;2} \cdot \widehat{(-\mathbf{k})})
= \sin(\theta_{1}) \sin(\pi - \theta_{2})
= \sin(\theta_{1}) \sin(\theta_{2})
= \sqrt{1 - \mu_{1}^{2}}\sqrt{1 - \mu_{2}^{2}}
\end{dmath}
where $\mu_{\alpha} \equiv \hatbf{k}_{\alpha} \cdot \hatbf{k}$ for $\alpha$= 1, 2. Thus, the usual term's expression is 
\begin{align}\label{usual}
    & \Bperp^{\mathrm{usual}}(\mathbf{k_{1}}, \mathbf{k_{2}}, \mathbf{k_{3}}) \nonumber \\
    & = \int \frac{d^{3}k}{\left( 2\pi \right) ^{3}} \left[\sqrt{1 - \mu_{1}^{2}}\sqrt{1 - \mu_{2}^{2}} \right]   P_{vv}(k) B_{\delta\delta\delta }( \mathbf{k_{1}}-\mathbf{k}, \mathbf{k_{2}}+\mathbf{k},\mathbf{k_{3}})
\end{align}

\subsection{The three extra terms}
\begin{figure}[!htb]
\center
{
\includegraphics[scale=0.57]{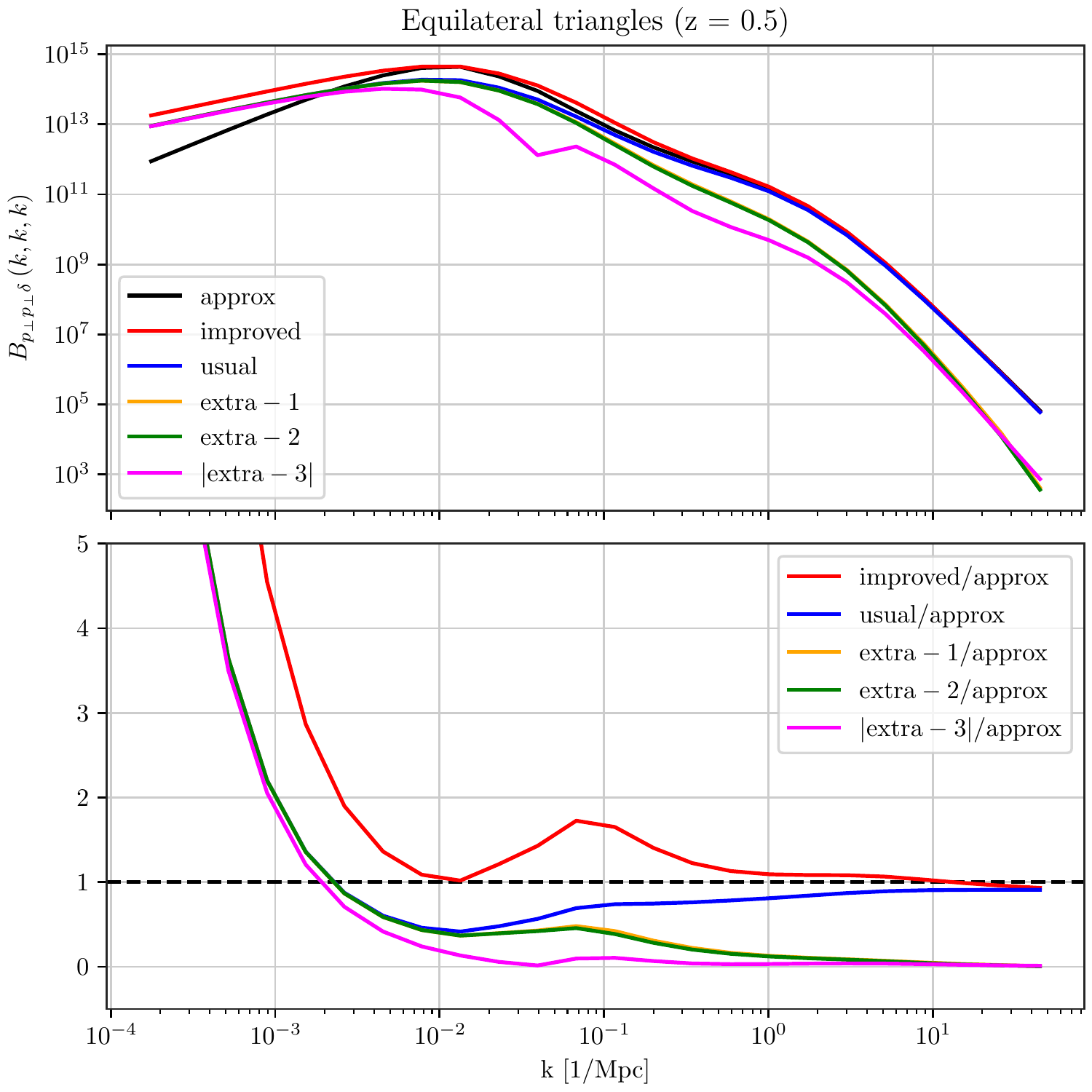} 
}
\caption{The hybrid bispectrum $\Bperp$ at $z$ = 0.5 for equilateral triangles of side lengths $k$. (top):\,The 4 terms as predicted by our theoretical model and their sum - the `improved' $\Bperp$ (red), compared with the approximate D05 model's prediction (black). (bottom):\,Ratios of these 4 terms and the improved $\Bperp$ with respect to the approximate $\Bperp$. In this equilateral case, extra-1 and extra-2 terms match.}
\label{fig:equi}
\end{figure}
We now calculate the three additional non-zero terms given in Table \ref{tab:wicks}, referring to each of them as the `extra$-i$' terms, because we find their geometric scalings to be {$\propto 1/k_i$}, for $i = 1, 2, 3$. For the extra$-1$ term, we have:
\begin{align}
    \langle \delta v\rangle \langle v \delta \delta \rangle & = \hatbf{p}_{\perp 1}^{i} \hatbf{p}_{\perp 2}^{j} \int \frac{d^{3}k}{\left( 2\pi \right) ^{3}}\int \frac{d^{3}k^{'}}{\left( 2\pi \right) ^{3}}  \nonumber \\
    & ~~~~~ \langle \delta( \mathbf{k_{1}}-\mathbf{k}) v^{j}(\mathbf{k'}) \rangle 
     \langle  v^{i}(\mathbf{k}) \delta( \mathbf{k_{2}}-\mathbf{k{'}}) \delta(\mathbf{k_{3}}) \rangle 
\end{align}

We simplify the Dirac delta functions analogous to the usual term's calculation and eliminate the integration variable $\mathbf{k}$. On substituting $\widehat{(\mathbf{k} + \mathbf{k}_{1})} =(\mathbf{k} + \mathbf{k}_{1} )/|\mathbf{k} + \mathbf{k}_{1}|$, we get
\begin{dmath} \label{extra-1}
  \Bperp^{\mathrm{extra-1}}(\mathbf{k_{1}}, \mathbf{k_{2}}, \mathbf{k_{3}}) \\
  = \int \frac{d^{3}k'}{\left( 2\pi \right) ^{3}} \left[\sqrt{1 - \mu_{1}^{2}}\sqrt{1 - \mu_{2}^{2}} \frac{k'}{|\mathbf{k'}+\mathbf{k}_{1}|}\right] \\
  P_{\delta v}(k') 
  B_{v\delta\delta }( \mathbf{k_{1}}+\mathbf{k'}, \mathbf{k_{2}}-\mathbf{k'},\mathbf{k_{3}}),
\end{dmath}
where $\mu_{\alpha} \equiv \hatbf{k}_{\alpha} \cdot \hatbf{k'}$ for $\alpha$ = 1, 2.

Similarly, we eliminate the integration variable $\mathbf{k'}$ to obtain the extra-2 term:
\begin{dmath} \label{extra-2}
  \Bperp^{\mathrm{extra-2}}(\mathbf{k_{1}}, \mathbf{k_{2}}, \mathbf{k_{3}}) \\
  = \int \frac{d^{3}k}{\left( 2\pi \right) ^{3}} \left[\sqrt{1 - \mu_{1}^{2}}\sqrt{1 - \mu_{2}^{2}} \frac{k}{|\mathbf{k}+\mathbf{k}_{2}|}\right] \\
  P_{v\delta}(k) 
  B_{\delta v \delta }( \mathbf{k_{1}}-\mathbf{k}, \mathbf{k_{2}}+\mathbf{k},\mathbf{k_{3}}),
\end{dmath}
where $\mu_{\alpha} \equiv \hatbf{k}_{\alpha} \cdot \hatbf{k}$ for $\alpha$ = 1, 2. As expected, the  expressions for the extra$-1$ and extra$-2$ terms above are symmetric in $\mathbf{k_{1}}$ and $\mathbf{k_{2}}$.

\begin{figure*}[t]
    \center
    {
    \includegraphics[width=0.70\textwidth]{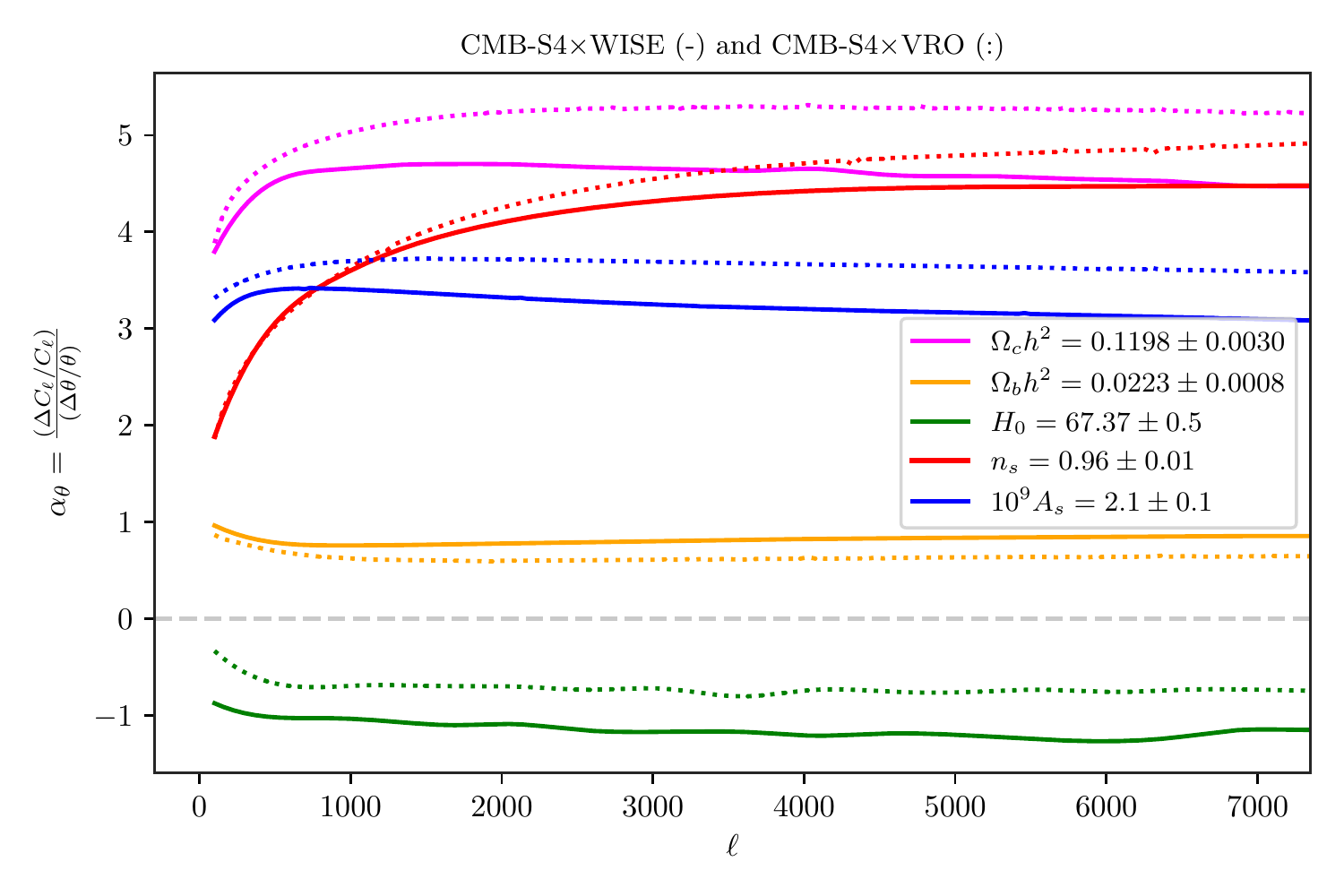}
    }
    \caption{Power-law indices $\alpha_{\theta}$ for each of the $\Lambda$CDM parameters $\pmb{\theta} = \{H_{0}, \Omega_{b}h^{2}, \Omega_{c}h^{2}, 10^{9}A_{s}, n_{s}\}$, showing the approximate dependence of $\CTTg$ on them as a function of $\ell$ for CMB-S4$\times$WISE (solid lines) and CMB-S4$\times$VRO (dotted lines).}
    \label{fig:alphas}
\end{figure*}

The extra$-3$ term looks different from them, since $\mathbf{k_{3}}$ corresponds to $\delta_{g}$, and not a momentum. We do an analogous calculation, eliminating the integration variable $\mathbf{k'}$, and substituting $\widehat{(\mathbf{k} + \mathbf{k}_{3})} =(\mathbf{k} + \mathbf{k}_{3} )/|\mathbf{k} + \mathbf{k}_{3}|$. Since $\mathbf{k_{3}} = -\mathbf{k_{1}}-\mathbf{k_{2}}$, 
\begin{align}\label{extra-3}
    & \Bperp^{\mathrm{extra-3}}(\mathbf{k_{1}}, \mathbf{k_{2}}, \mathbf{k_{3}}) \nonumber \\
    & = \int \frac{d^{3}k}{\left( 2\pi \right) ^{3}} 
  \bigg[- \sqrt{1 - \mu_{1}^{2}}\sqrt{1 - \mu_{2}^{2}} \frac{k}{|\mathbf{k}+\mathbf{k}_{3}|} + \nonumber\\ 
  & ~~~~~~~~~~~~~~~~~~ \sqrt{1 - \mu_{1}^{2}}\sqrt{1 - \mu_{12}^{2}}\frac{k_{1}}{|\mathbf{k}+\mathbf{k}_{3}|}\bigg] 
  \nonumber \\
  & ~~~~~~~~~  P_{\delta\delta}(|k-k_{1}|)
  B_{vv \delta }( \mathbf{k}, \mathbf{k_{1}}+\mathbf{k_{2}}-\mathbf{k},\mathbf{k_{3}}),
\end{align}
where $\mu_{\alpha} \equiv \hatbf{k}_{\alpha} \cdot \hatbf{k}$ for $\alpha$ = 1, 2, and $\mu_{12} \equiv \hatbf{k}_{1} \cdot \hatbf{k}_{2}$. Note that this extra$-3$ term is generally negative at large scales and positive at small scales since it has two parts. A similar expression is alternatively obtained if we choose to eliminate the integration variable $\mathbf{k}$, where the roles of $k_1$ and $k_2$ get exchanged. 

Thus, the `improved' hybrid bispectrum $\left(\Blos = \frac{1}{2} \Bperp\right)$ predicted by our derived improved model is:
\begin{equation} \label{full}
\Blos = (1/2) \left[\Bperp^{\mathrm{usual}}+ \Bperp^{\mathrm{extra-1}} + \Bperp^{\mathrm{extra-2}} + \Bperp^{\mathrm{extra-3}} \right].
\end{equation}

In Figure \ref{fig:equi} (top panel), we show the individual contributions of the four terms of $\Bperp$ (plotting the absolute value of extra$-3$) and their sum - the improved model's prediction for $\Bperp$ (red), for the special case of equilateral triangles at $z = 0.5$. We also show the approximate D05 model's prediction (black). The bottom panel shows the ratios of these four terms and of their sum with respect to the approximate prediction. For this particular case, the models match reasonably at the smallest scales but differ at larger scales. When the total estimator $\CTTg$ is computed, the CMB filters applied suppress $\Bperp$ at large scales (small $k$), and $\Bperp$ contributions from all triangle shapes are summed together. 

\section{Understanding the Cosmological Dependence} \label{cosmodep}
In Section \ref{cosmo-dep}, we show that the marginalized uncertainty on $\Ak$ goes from a sub-percent level to $\sim7\%$ when cosmology is allowed to vary. To understand this result roughly, we approximately model the dependence of $\CTTg$ (denoted here as `$C_{\ell}$') on a parameter $\theta$ as a power-law scaling: $C_{\ell}\propto \theta^{\alpha_{\theta}}$. We estimate the index $\alpha_{\theta}$ for each of the $\Lambda$CDM parameters as: $\alpha_{\theta}\approx \frac{(\Delta C_{\ell}/C_{\ell})}{(\Delta\theta/\theta)}$, where $\Delta\theta$ are the same step-sizes used in Section \ref{cosmo-dep} for computing partial derivatives, and $\Delta C_{\ell}$ are the corresponding changes in the signal. 

From Eq.\eqref{eq-scaling}, since $P_{\delta\delta}\propto A_s$, we expect that $C_{\ell}$ roughly scales as $A_s^{3}$, as seen in Figure \ref{fig:alphas} across scales. $P_{\delta\delta}$ depends non-linearly on the total matter density today, $\Omega_{m}$, while the linear growth rate $f$ roughly scales as $\Omega_{m}^{0.55}$. Since $\Omega_{m}=\Omega_{c}+\Omega_{b}$, and the fiducial value of $\Omega_{c}$ is about 5 times $\Omega_{b}$, $C_{\ell}$ has the strongest dependence on $\Omega_{c}h^{2}$ with an $\alpha_{\theta}$ of $\sim4.5$-5, while $\alpha_{\Omega_{b}h^{2}}\approx 1$. Because $n_s$ itself is the slope of the primordial spectrum, its $\alpha_{\theta}$ varies strongly with $\ell$ and $C_{\ell}$'s dependence on it is not well-modeled as a power-law. 

Given rough power-law scalings, each $\theta$ with a residual error of $\sigma(\theta)$ contributes $\sim(|\alpha_{\theta}|\sigma(\theta)/\theta)$ to the relative propagated error on $C_{\ell}$. Thus, if they were fully uncorrelated, $A_s$ and $\Omega_{c}h^{2}$ together would imply a propagated error of $\sim9\%$ on $\Ak$. While the actual uncertainty on $\Ak$ is lower due to degeneracies between $\Lambda$CDM parameters, a majority of it arises from residual uncertainties on $A_s$ and $\Omega_{c}h^{2}$ in the \textit{Planck} prior. The dependence on $\Lambda$CDM parameters of the $\CTTg$ signal from SO$\times$WISE is similar to that from CMB-S4$\times$WISE, although it differs for the signal from CMB-S4$\times$VRO due to the $z$-dependent effects of cosmology. We discuss this further in Section \ref{fish-mnu}.

\bibliography{library}
\end{document}